\documentclass[10pt]{iopart}

\expandafter\let\csname equation*\endcsname\relax
\expandafter\let\csname endequation*\endcsname\relax

\usepackage[numbers,sort&compress]{natbib}

\usepackage[justification=justified]{caption}
\usepackage{multicol, blindtext}
\usepackage{amssymb}

\usepackage{dsfont}
\usepackage{color} 
\usepackage[framemethod=TikZ]{mdframed}
\usepackage{algpseudocode}
\usepackage{etoolbox}
\apptocmd{\sloppy}{\hbadness 10000\relax}{}{}

\usepackage{textcomp}

\usepackage{url}

\usetikzlibrary{positioning}
\usepackage[breaklinks=true,colorlinks=true,linkcolor=blue,urlcolor=blue,citecolor=blue]{hyperref}
\usepackage{physics}

\usepackage[utf8]{inputenc}

\usepackage{graphicx}
\usepackage{subcaption}

\usepackage{iopams}

\usepackage{pstricks}
\usepackage{float}
\usepackage[T1]{fontenc}
\usepackage{bbm}
\usepackage{dsfont}
\usepackage[linesnumbered,ruled,vlined]{algorithm2e}
\SetKwInput{kwInit}{Init}
\usepackage{mathtools}

\usepackage{xcolor}
\usepackage{siunitx}
\usepackage{pifont} 

\newcommand{\newblock}{}

\DeclareUnicodeCharacter{2212}{-}

\newcommand{\paramtheta}{\boldsymbol{\theta}}


\newcommand{\computerfont}[1]{{\fontfamily{cmtt}\selectfont #1}}

\newcommand{\x}{\boldsymbol{x}}
\newcommand{\y}{\boldsymbol{y}}
\newcommand{\vbs}{\boldsymbol{v}}

\def\andname{\hspace*{-0.5em}}

\def\orcid#1{\kern -0.4em\href{https://orcid.org/#1}{\includegraphics[keepaspectratio,width=0.7em]{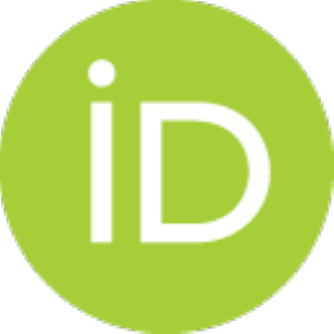}}}

\newcommand{\Ansatze}{\text{Ans\"{a}tze}}

\newcommand{\ansatze}{\text{ans\"{a}tze}}

\begin{document}

\title{Quantum versus Classical Generative Modelling in Finance}

\author{Brian Coyle${^{*, \mathsection}}$, Maxwell Henderson${^\dagger}$, Justin Chan Jin Le${^\dagger}$, Niraj Kumar${^*}$, Marco Paini${^\dagger}$, Elham Kashefi${^{*, \ddagger}}$}

\address{${^*}$School of Informatics, 10 Crichton Street, Edinburgh, United Kingdom, EH8 9AB.}
\address{${^\dagger}$Rigetti Computing.}
\address{${^{\ddagger}}$CNRS, LIP6, Sorbonne Universit\'{e}, 4 place Jussieu, 75005 Paris, France.}
\ead{${^{\mathsection}}$brian.coyle@ed.ac.uk}
\vspace{10pt}

\begin{abstract}
Finding a concrete use case for quantum computers in the near term is still an open question, with machine learning typically touted as one of the first fields which will be impacted by quantum technologies. In this work, we investigate and compare the capabilities of quantum versus classical models for the task of generative modelling in machine learning. We use a real world financial dataset consisting of correlated currency pairs and compare two models in their ability to learn the resulting distribution - a restricted Boltzmann machine, and a quantum circuit Born machine. We provide extensive numerical results indicating that the simulated Born machine always at least matches the performance of the Boltzmann machine in this task, and demonstrates superior performance as the model scales. We perform experiments on both simulated and physical quantum chips using the Rigetti forest platform, and also are able to partially train the largest instance to date of a quantum circuit Born machine on quantum hardware. Finally, by studying the entanglement capacity of the training Born machines, we find that entanglement typically plays a role in the problem instances which demonstrate an advantage over the Boltzmann machine.

\end{abstract}

%
\vspace{1pc}
\ioptwocol
%

\section{Introduction} \label{sec:intro}

The prediction power of machine learning algorithms is limited by the quality of the datasets used to train the models. In the age of big data, possessing high-quality data can offer significant competitive advantage to institutions who utilize machine learning in their core business operations such as Facebook, Google and Amazon. However, for many organizations high-quality data can be scarce. This is because training data for industrial problems are often plagued by erroneous information, limited by privacy and over-fitting. Hence, high-quality data can be expensive or even impossible to obtain especially for machine learning applications at industrial scales. Synthetic data generation (SDG) bridges the gap for training better machine learning models when such data is not readily available. Rather than collecting raw data, SDG uses statistical methods, simulation modeling, and neural networks to generate a synthetic equivalent of the real-world data set (i.e.\@ sample generation). SDG allows users to overcome data scarcity, avoid privacy issues, and overcome over-fitting problems at lower costs. This is achieved by SDG removing erroneous or mislabeled data, as each sample is generated from predefined parameters to produce clean and machine learning-ready datasets. SDG can also produce realistic data for unobserved scenarios to train more generalized models. In machine learning terms, SDG is typically achieved by \emph{generative modelling} or distribution learning. Using quantum models for SDG has garnered interested due to the ease of generating data samples (alternatively, performing the `inference' step) from a quantum distribution, whereas even the very act of sample generation can be difficult classically, which we elaborate on through the text.

In terms of quantum capabilities, we are now firmly in the noisy intermediate scale quantum (NISQ)~\cite{preskill_quantum_2018} era, where we have access to small, error-prone quantum computers, but which are sufficiently powerful to be able to address problems which are not classically simulatable~\cite{arute_quantum_2019}. However, finding a useful application for such devices is a non-trivial task, with quantum chemistry~\cite{mcardle_quantum_2020, arute_hartree-fock_2020} or quantum optimization~\cite{farhi_quantum_2014, arute_quantum_2020} being the usual suspects for areas in which to search. Problems in finance have also proved to be a lucrative area of study, \cite{zoufal_quantum_2019, ramos-calderer_quantum_2019, rebentrost_quantum_2018, mugel_dynamic_2020}. With each discovered use case, an argument is frequently required as to why such a problem could not have been tackled by purely classical methods. The primary approaches to gain an advantage with quantum computers study the computational time complexity in solving these problems. The claims of exponential speedups~\cite{shor_polynomial-time_1997, harrow_quantum_2009} in these cases usually rely on the non-existence of unlikely relationships between computational complexity classes. However, simply solving the problem faster is not the only way in which quantum computers can gain victories. Alternatively, one can examine other relevant problem dimensions, such as accuracy of solution, which is the goal we aim for in this work.

We explore two different machine learning approaches for generating synthetic financial market data. One model is completely classical (although trained using simulated quantum methods): the restricted Boltzmann machines (RBM) and the other is completely quantum in nature: a quantum circuit Born machines (QCBM). This is similar to other recent works~\cite{kondratyev_non-differentiable_2020}, which addressed financial problems with these two models and found that the Born machine has the capacity to outperform the Boltzmann machine, when it comes to generating synthetic data. In this work, we draw a similar conclusion by enforcing similar constraints on both models in order to draw a fair comparison.

In \Sref{sec:generative_modelling} we discuss the main ideas involved in generative modelling, and elaborate on the two models we use for this task. We also discuss the financial dataset we use for training. In \Sref{sec:model_structures}, we detail the specific architectures for the Boltzmann and Born machines, namely the underlying graph structures and the circuit $\Ansatze$ for the QCBM. In \Sref{sec:training_procedures}, we describe the training protocols we use for each model and finally in \Sref{sec:results} we detail the numerical results we find, and showcase examples where the Born machine outperforms the Boltzmann machine in learning the financial dataset. We present simulated and experimental results implemented on the Rigetti QPU \cite{karalekas_quantum-classical_2020} using Quantum Cloud Services (QCS\texttrademark). Finally, we conclude in \Sref{sec:discussion} and discuss future work.

\begin{figure}[t]
    \centering
\includegraphics[width=0.8\columnwidth, height=0.7\columnwidth]{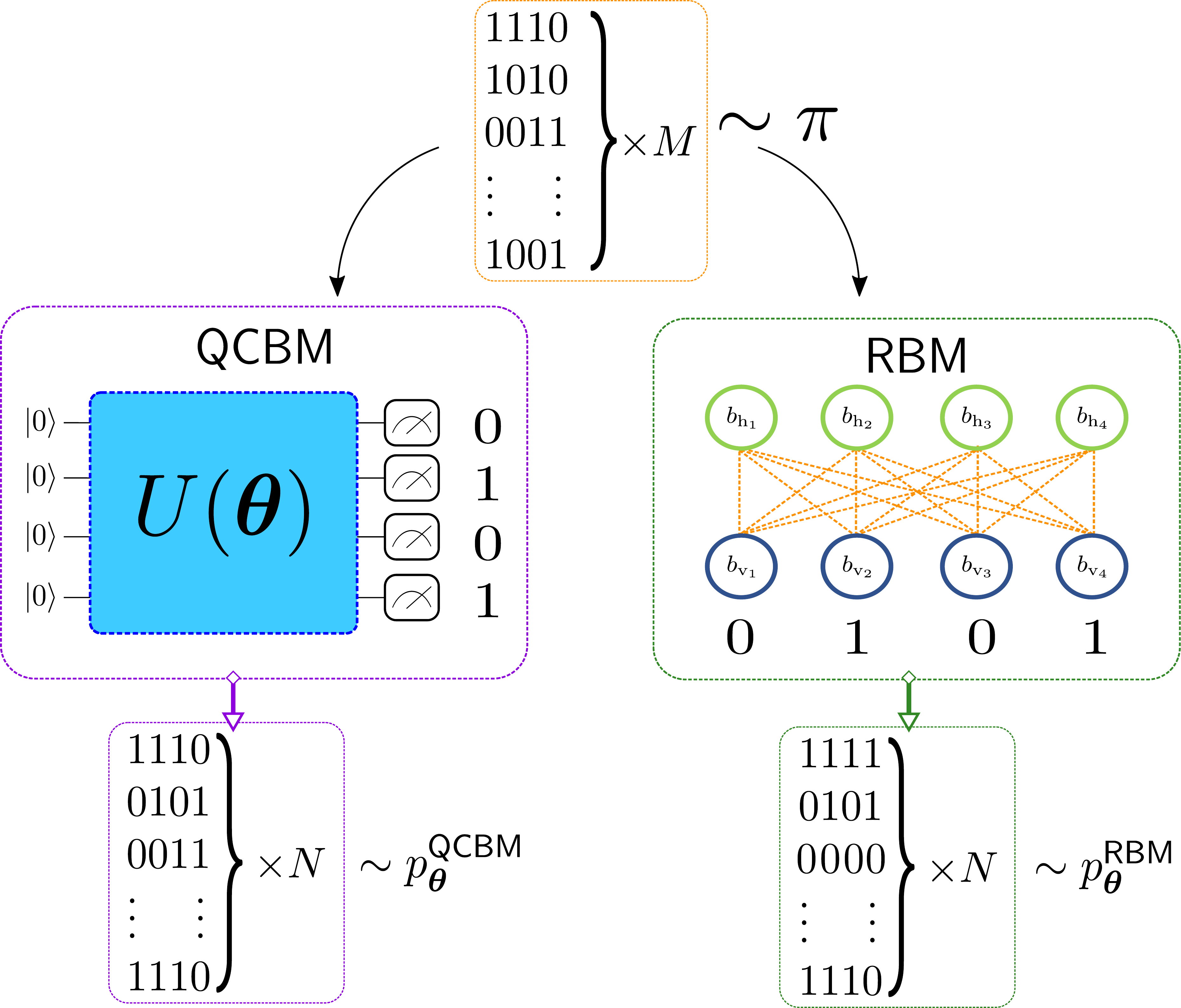}
    \caption{Illustration of synthetic data generation by a quantum circuit Born machine (QCBM) versus a restricted Boltzmann machine (RBM) for binary strings of length $4$. Generative modelling involves learning a representation of an underlying distribution ($\pi$) from $M$ samples. The trained model can generate $N$ samples where $N$ can be larger than $M$. The samples associated to the QCBM are the binary results from measuring qubits in (typically) the computational basis and generate $p^{\mathsf{QCBM}}_{\paramtheta}$, whereas for the RBM, they are associated to configurations of the visible nodes, $p^{\mathsf{RBM}}_{\paramtheta}$. We judge the quality of the samples by the similarity of the generated distributions to $\pi$.}
    \label{fig:born_versus_boltzmann_illustration}
\end{figure}

\section{Generative Modelling}\label{sec:generative_modelling}

Generative models are powerful machine learning models, which essentially aim to learn a probability distribution, denoted $\pi$, over some data (say vectors, $\x$), which is sampled from $\pi$, $\boldsymbol{x}\sim \pi(\x)$. A typical use case is in classification tasks, where a generative model seeks to learn the joint distribution over data and labels, $y$, $\pi(\x, y)$.  We assume the distribution in question is defined over the space of binary strings of length $n$, $\{0, 1\}^n$. A generative model can be typically parameterised by some parameters, $\paramtheta$, and are represented by an output `model' distribution over the data, which is a function of those parameters, $p_{\paramtheta}(\x)$. The goal of training a generative model is to force the model distribution as close as possible to the data distribution, relative to some measure. This is done by finding a suitable setting of the parameters, typically using some optimization routine. In practice however, we typically do not have access to the true distributions (meaning their explicit probability density functions or otherwise) This is inevitably true when using implicit models
\footnote{These are models for which we do not have explicit access to the underlying probability density function \cite{mohamed_learning_2016}.}
like generative adversarial networks (GANs)~\cite{goodfellow_generative_2014} or quantum circuit distributions, which by their very nature are distributions which are not directly accessible~\cite{arute_quantum_2019} due to the classical intractability of them. In this work, we assume we have $N$ samples from the model distribution, $\{\x_i\}_{i=1}^N, \x_i \sim p_{\paramtheta}(\x)$ and $M$ samples from the data distribution, $\{\y_j\}_{j=1}^M, \y_j \sim \pi(\y)$.

Common use cases for generative models are in image generation, but they have also received interest from the quantum computing community, as the acceleration in training of generative models using quantum techniques was one of the early areas of interest in the field of quantum machine learning~\cite{wiebe_quantum_2015}. The focus of the area has shifted somewhat in recent years, from accelerating training and inference of \emph{classical} models using quantum techniques, to the development of \emph{completely new} models in the quantum world. One of the earliest examples of which is the \emph{quantum Boltzmann machine} (QBM), which is a generalization of the classical Boltzmann machine (see \Sref{ssec:boltz_machine}). This was followed by the introduction of Born machines~\cite{cheng_information_2018} and \emph{quantum circuit Born machines} (QCBMs)~\cite{benedetti_generative_2019, liu_differentiable_2018}, which sample from the fundamentally quantum distribution underlying a pure state of a quantum system. One of the most recent additions to this family are Hamiltonian based models and the variational quantum thermalizer (VQT)~\cite{verdon_quantum_2019}, which generalizes all of the above since it contains the distribution provided by a \emph{mixed} quantum state as the underlying model. The latter is also a generalization of `energy-based' models, of which the Boltzmann machine is an example. Furthermore, quantum generative models are some of the most promising applications for near term quantum computers since their nature aligns them closely with demonstrations of `\emph{quantum supremacy}'~\cite{arute_quantum_2019} and such connections have recently been made~\cite{ coyle_born_2020, sweke_quantum_2020} with extensions into different architectures~\cite{tangpanitanon_expressibility_2020}.

In this work we focus on two of these models in order to make a direct comparison and study any potential indication of quantum advantage for these models over purely classical generative models. We investigate a Born machine and a restricted Boltzmann machine (RBM) and make a thorough comparison between the two for a generative modelling task. We do this using a realistic dataset in a financial application, which facilitates a simple way to compare the models at differing scales. Our motivation is the work of \cite{kondratyev_market_2019} which showed the outperformance of an RBM over parametric models, for this dataset, which are the common tool used in the finance industry. This was subsequently followed by the subsequent outperformance of the RBM by a Born machine~\cite{kondratyev_non-differentiable_2020} on the same dataset. However, the degree to which this advantage was observable was not obvious. This research and \cite{kondratyev_non-differentiable_2020} supplements the work of~\cite{alcazar_classical_2020} which demonstrated a similar outperformance of an RBM by a QCBM, but for a different problem domain. Our work expands on the latter by running larger problem instances on simulators and physical hardware, using alternative training methods, and also using alternative methods of comparison of the models. Finally, drawing a comparison between a Born and Boltzmann machine is part of the goal of \cite{cheng_information_2018}, in which they consider the problem from a mutual information point of view. They further conjecture that properties such as mutual information of the dataset, and entanglement entropy in the target problem, an/or model would be useful in determining problems where the Born machine could have superior performance over an RBM.


\subsection{Born Machine}\label{ssec:born_machine}

A Born machine~\cite{cheng_information_2018} is a fundamentally quantum model, which achieves synthetic data generation by generating samples according to Born's rule of quantum mechanics. The fundamentally non-classical nature of the model has provided motivation for why it can outperform classical models in at least its expressive power~\cite{coyle_born_2020}. This expressive power translates in an ability to represent certain distributions efficiently which cannot be done by any classical model, for example, those utilized in a recent demonstration of quantum computational supremacy~\cite{arute_quantum_2019}.   

In the most common scenario, a binary sample, $\x \in \{0, 1\}^n$, is generated from a quantum state, $\rho$, according to:
\begin{equation}\label{eqn:born_rule}
    \x \sim p(\x) = \Tr\left(\ketbra{\x}{\x}\rho\right)
\end{equation}
where $\ketbra{\x}{\x}$ is the projector onto the computational basis state described by $\x$. In order to obtain a trainable machine learning model, we parameterize the state: $\rho \rightarrow \rho_{\paramtheta}$. We also further consider the scenario where the parameterised state is a pure state, i.e.\@ $\rho_{\paramtheta} := \ketbra{\psi_{\paramtheta}}{\psi_{\paramtheta}}$. In this case, the correlations present in the model will be of a purely quantum nature. The parameterised distribution is then:
\begin{equation} \label{eqn:parameterised_qcbm}
 \x \sim  p^{\mathsf{QCBM}}_{\paramtheta}(\x) =  |\braket{\x}{\psi_{\paramtheta}}|^2
\end{equation}
Finally, if the state, $\ket{\psi_{\paramtheta}}$ is generated by a quantum circuit (as opposed to, for example, by a continuous time Hamiltonian evolution), the model is referred to as a \emph{quantum circuit} Born machine \cite{liu_differentiable_2018, benedetti_generative_2019} (QCBM). In this form, the ease of performing inference becomes apparent: once trained, the parameterized quantum state prepared by a quantum circuit and then simply measured. The measurement results then constitute an (approximate) sample from the data distribution. Furthermore, utilizing quantum randomness as a sample generation mechanism this way relaxes the need to input randomness into the model as is usually done to build GANs. However, we mention that inputting randomness has been considered in the quantum case~\cite{romero_variational_2019} as well, although the advantage of doing so has yet to be explored.

A generalization of the above can be achieved by relaxing the purity assumption of the underlying state, and doing so results in quantum Hamiltonian based models~\cite{verdon_quantum_2019}, which instead can carry \emph{both} classical and quantum correlations.

In order to find a good fit to the data distribution, $\pi$, such that the model, $p_{\paramtheta}(\x)$, can effectively generate synthetic data, an optimization routine is invoked to search over the space of possible states $\ket{\psi_{\paramtheta}}$. Since Born machines are implicit models, careful consideration must be given to the choice of optimization routine, since any optimizer must be able to effectively, and efficiently, deal with samples alone. One may consider quantum training procedures~\cite{verdon_universal_2018}, but more commonly the optimization procedure will be a fully classical routine. This makes these models hybrid quantum-classical in nature and therefore friendly to NISQ devices, only using the quantum resource when necessary.


\subsection{Boltzmann Machine}\label{ssec:boltz_machine}

\begin{figure}[t]
    \centering
    \includegraphics[width=\columnwidth]{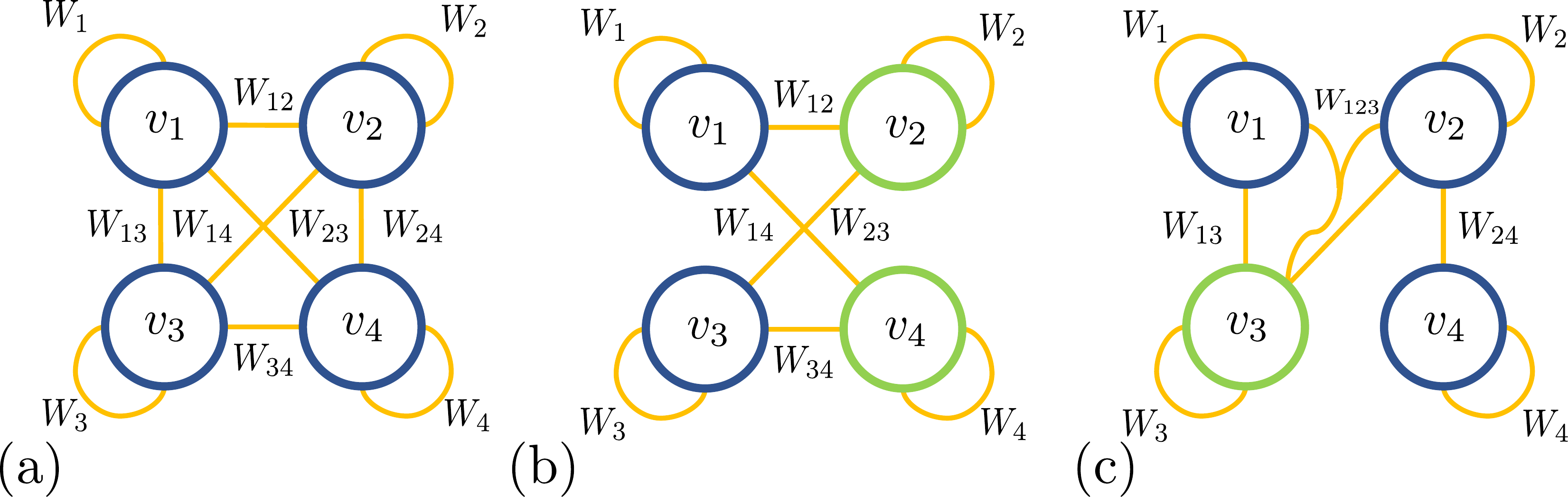}
    \caption{Visualizing various possible GBM architectures for a network with 4 nodes, such as a (a) fully-connected visible network, (b) restricted Boltzmann machine (RBM), and (c) partially-connected higher order Boltzmann machine.  All visible nodes are shown in blue and hidden nodes are shown in green.}
    \label{fig:gbm_architecture_examples}
\end{figure}

Generalized Boltzmann machines (GBMs) are graphical models with powerful synthetic data-generation capabilities. While GBMs can vary significantly in terms of how they are applied to various problems and their particular architectures (see some example architectures in Figure \ref{fig:gbm_architecture_examples}), they all share some defining characteristics. The model architecture is defined by a graph $\mathsf{G}$, which consists of a set of edges, $\mathcal{E}$, and nodes (vertices) which we denote $\mathcal{N}$. Each edge, $e \in \mathcal{E}$ has a corresponding edge \emph{weight}, $W_e \in \mathcal{W}$. In generality, the edge weights can also be self-loops (biases in standard Boltzmann machine terminology), or hyper-edges (edges connecting more than two nodes) as illustrated in \Fref{fig:gbm_architecture_examples}(c). Crucially, the nodes are typically partitioned into \emph{visible} and \emph{hidden} nodes, $\mathcal{N} = \mathcal{N}_{\textrm{v}} \bigcup \mathcal{N}_{\textrm{h}}$. The visible nodes directly model some aspect of the data distribution, while the hidden nodes are used for capturing features of the data, and are not tied to any particular aspect of it. As such the hidden nodes typically correspond directly to the expressive power of the model. Finally, a sample generated by the GBM is distributed according to the Boltzmann distribution:
\begin{equation} \label{eqn:gbm_probability}
    \x \sim p^{\mathsf{GBM}}_{\paramtheta}(\x) = \frac{e^{-\beta E(\x)}}{\mathcal{Z}},
\end{equation}
$p^{\mathsf{GBM}}_{\paramtheta}(\x)$ is the probability to observe the visible nodes in some state $\boldsymbol{n}_\mathrm{v} = \x$, and describes the model distribution for the Boltzmann machine. $\boldsymbol{n}_\mathrm{v}$ is a particular state (corresponding to a binary vector) of the visible nodes in $\mathcal{N}_{\mathrm{v}}$. In this case, the model parameters are the weights of the machine, $\paramtheta = \mathcal{W}$. $E$ and $\mathcal{Z}$ is the model \emph{energy} (defining an energy based model~\cite{goodfellow_deep_2016}) and \emph{partition function} respectively, and are defined by:
\begin{align}
    E(\vbs) &:= -\sum_{e \in \mathcal{E}} W_{e} \prod_{v_{i} \in e} v_{i},\label{eqn:gbm_energy} \\
    \mathcal{Z} &:= \sum_{\vbs}{e^{-\beta E(\vbs)}} \label{eqn:partition_function}
\end{align}
The notation, $v_{i} \in e$, refers to the nodes connected to edge $e$, and $\beta$ is an effective inverse temperature term. The sum in \Eref{eqn:partition_function} is taken over all possible binary vectors $\vbs \in \{0, 1\}^n$.
In this work, we focus specifically on the \emph{restricted} version of the Boltzmann machine (RBM) corresponding to \Fref{fig:gbm_architecture_examples}(b), and we discuss this specification further in \Sref{ssec:boltzmann_machine_structure}. In this case, we denote $p^{\mathsf{GBM}}_{\paramtheta}(\x) \rightarrow p^{\mathsf{RBM}}_{\paramtheta}(\x)$ where the latter distribution is generated by marginalizing over hidden units.

Finally, while all of the above is purely classical (in contrast to the Born machine, a GBM carries only classical correlations), the extension of the \emph{model itself} into the quantum world has also been proposed in the quantum Boltzmann machine ~\cite{amin_quantum_2018, kieferova_tomography_2017, song_geometry_2019, wiebe_generative_2019} as we mentioned above. In this framework, the energy function, \Eref{eqn:gbm_energy} is replaced by a quantum Hamiltonian and the model distribution in question is generated by sampling from the thermal state of this Hamiltonian, mimicking a Boltzmann distribution. This thermal state can be prepared either by quantum annealing~\cite{amin_quantum_2018} or by a gate based approach~\cite{verdon_quantum_2017}. By introducing off-diagonal terms in this Hamiltonian, non-trivial quantum behavior can be exploited, and the model inherits some characteristics of a Born machine (i.e.\@ some of the randomness originates from Born's rule). 

In this work, however, we focus on the GBM as a completely classical object, which we detail in \Sref{sec:model_structures}, however, we do leverage quantum inspired training methods which are discussed in \Sref{sec:training_procedures}. Furthermore, as mentioned we only study the RBM here, but we discuss the extension of the methods in this work to the more general Boltzmann machine structures in \Sref{sec:discussion}.


\subsection{A Financial Dataset} \label{ssec:financial_data}

\begin{figure}[t]
    \centering
    \includegraphics[width=\columnwidth, height=0.3\columnwidth]{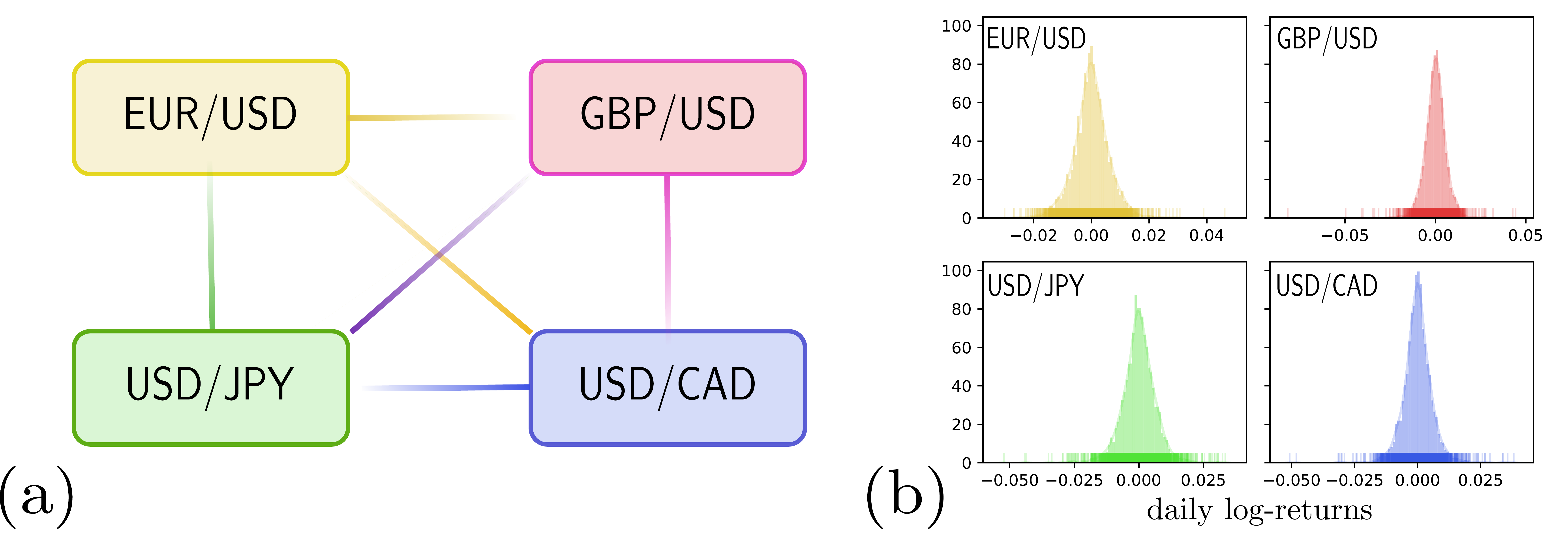}
    \caption{We use data generated from FX spot prices of the above currency pairs. The generative model aims to learn correlations between each pair based on a $16$ bit binary representation. (a) The selection  of currency pairs we use, and (b) the marginal distributions of the log-returns of each pair over a $20$ year period. We aim to learn the joint distribution of subsets of the pair in this work.}
    \label{fig:currency_pairs}
\end{figure}

In order to perform SDG, we require some dataset to learn. In this work, we focus on one of a financial origin, in particular one considered by \cite{kondratyev_market_2019}. This dataset comtains $5070$ samples of daily log-returns of $4$ currency pairs between $1999-2019$ (see \Fref{fig:currency_pairs}). In order to fit on the binary architecture of the Born and Boltzmann machines, the spot prices of each currency pair are converted to $16$ bit binary values, resulting in samples of $64$ bits long. This discretisation provides a convenient method for fitting various problem sizes onto models with different numbers of qubits or visible nodes for the Born machine or RBM respectively. In particular, we can tune both the number of currency pairs ($i$), and the precision of each pair ($j$) so the problem size is described by a tuple $(i, j)$. For example, as we revisit in \Sref{sec:model_structures}, a $12$ qubit Born machine can be tasked to learn the distribution of $4$ currency pairs at $3$ bits of precision, $3$ pairs with $4$ bits or $2$ pairs at $6$ bits of precision.

\section{Model Structures} \label{sec:model_structures}

Here we provide specific details about the model architectures we choose to use, in order to derive as fair a comparison as possible. In the first instance, we choose to only train the bias terms in the RBM (the self-loops in \Fref{fig:gbm_architecture_examples}) for simplicity. We also fix the number of parameters in the Born machine by the number of layers, and then match the number of parameters in the RBM to this, since it is simpler to grow the number of RBM parameters by simply adding extra nodes. 

\subsection{Born Machine Ansatz} \label{ssec:born_machine_strucure}

\begin{figure}[t]
    \centering
    \includegraphics[width=\columnwidth, height=0.25\columnwidth]{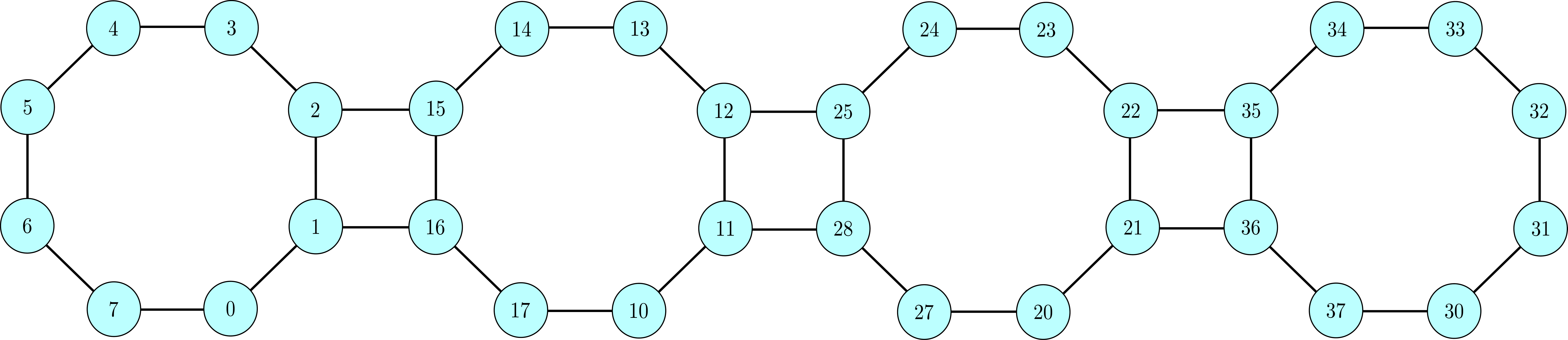}
    \caption{\computerfont{Aspen 7}/\computerfont{Aspen 8} $32$ qubit chip designs. Note not all connections shown above are directly accessible on the chip itself.}
    \label{fig:aspen_7_chip}
\end{figure}

The Ansatz which we use for the QCBM is hardware efficient as we endeavor to run the model on real quantum hardware. We also restrict the number of parameters in the circuit to match the number used in the RBM, following~\cite{alcazar_classical_2020}. We choose this hardware native approach to closely fit the structure of Rigetti's chip design (the structure of the \computerfont{Aspen-7} and \computerfont{Aspen-8} can be seen in \Fref{fig:aspen_7_chip}). Furthermore, we solely parameterize the single qubit unitaries to avoid compilation overheads arising out of two qubit unitary parameterization. If we were to do so, we could employ a similar strategy to \cite{schuld_circuit-centric_2020}, which uses `blocks' of parameterized unitaries in such a way to enforce a linear scaling of the number of parameters with the number of qubits, when building a quantum classifier.

\begin{figure}
        \centering
        \includegraphics[width=\columnwidth, height=0.4\textwidth]{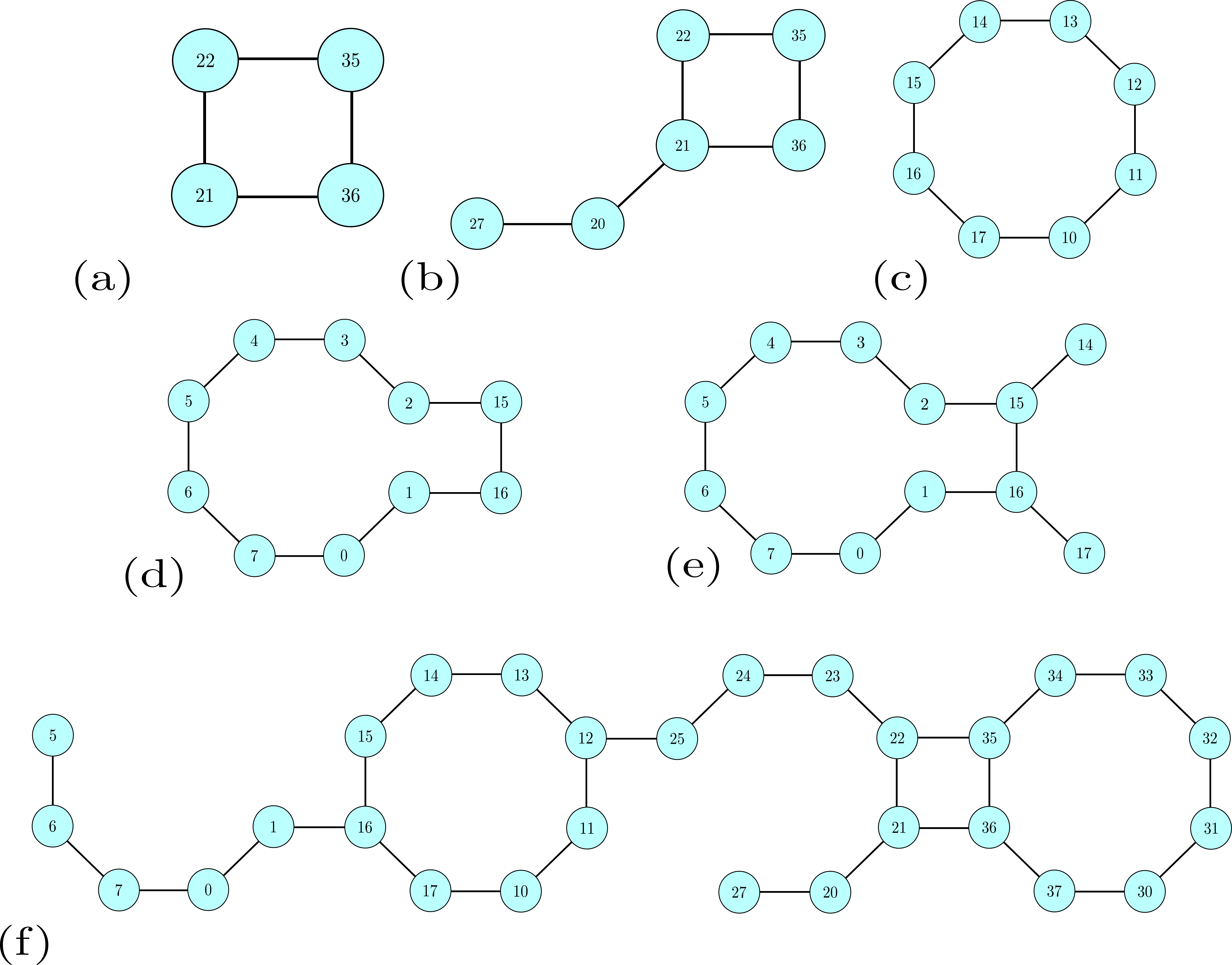}
    \caption{Select sublattices from the \computerfont{Aspen-7} and \computerfont{Aspen-8 } chips, corresponding to different problem sizes. Figure shows the (a) \computerfont{Aspen-7-4Q-C}, (b) \computerfont{Aspen-7-6Q-C}, (c) \computerfont{Aspen-7-8Q-C}, (d) 10 qubit \computerfont{Aspen-8}, (d) 12 qubit \computerfont{Aspen-8} and (f) \computerfont{Aspen-7-28Q-A} sublattices. Using these topologies we can fit problems of size $(2, 2), (2, 3), (2, 4), (2, 5), (2, 6)$ and $(4, 7)$ respectively, where the notation $(i,j)$ indicates $i$ currency pairs, each described by $j$ bits of precision.}
    \label{fig:aspen_sublattices}
\end{figure}

We run all experiments using the Rigetti \computerfont{Aspen 7} and \computerfont{Aspen 8} chips, which are designed to contain $32$ qubits, however some qubits are not available. Each QPU can be divided into sublattices containing fewer qubits, some examples can be seen in \Fref{fig:aspen_sublattices}.
The largest sublattice on the \computerfont{Aspen-7} chip is the \computerfont{Aspen-7-28Q-A} which contains $28$ usable qubits (seen in \Fref{fig:aspen_sublattices}(f)).

\begin{figure}[t]
    \centering
    \includegraphics[width=0.9\columnwidth, height=0.3\columnwidth]{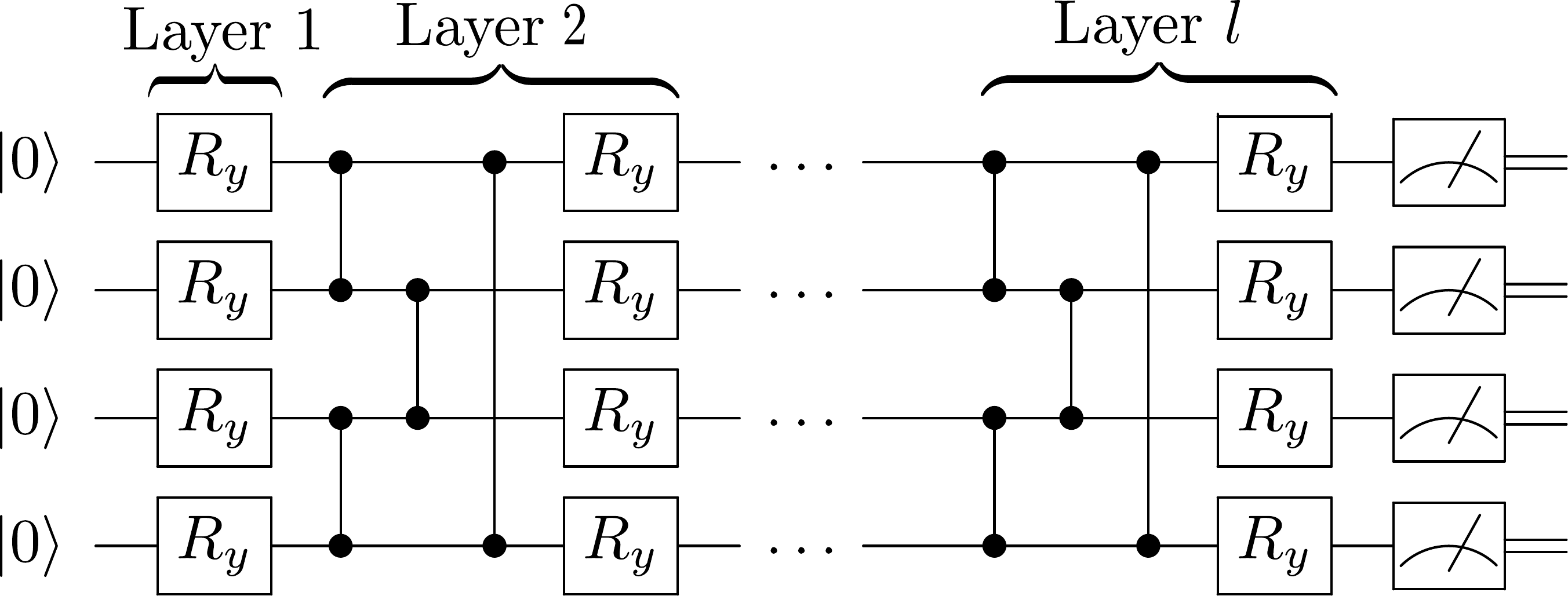}
    \caption{Hardware efficient circuit for the \computerfont{Aspen-7-4Q-D}, with $l$ layers using the native entanglement structure native to the chip.}
    \label{fig:aspen_7_4q_circuit_ansatz}
\end{figure}

\begin{figure}[t]
    \centering
    \includegraphics[width=\columnwidth, height=0.9\columnwidth]{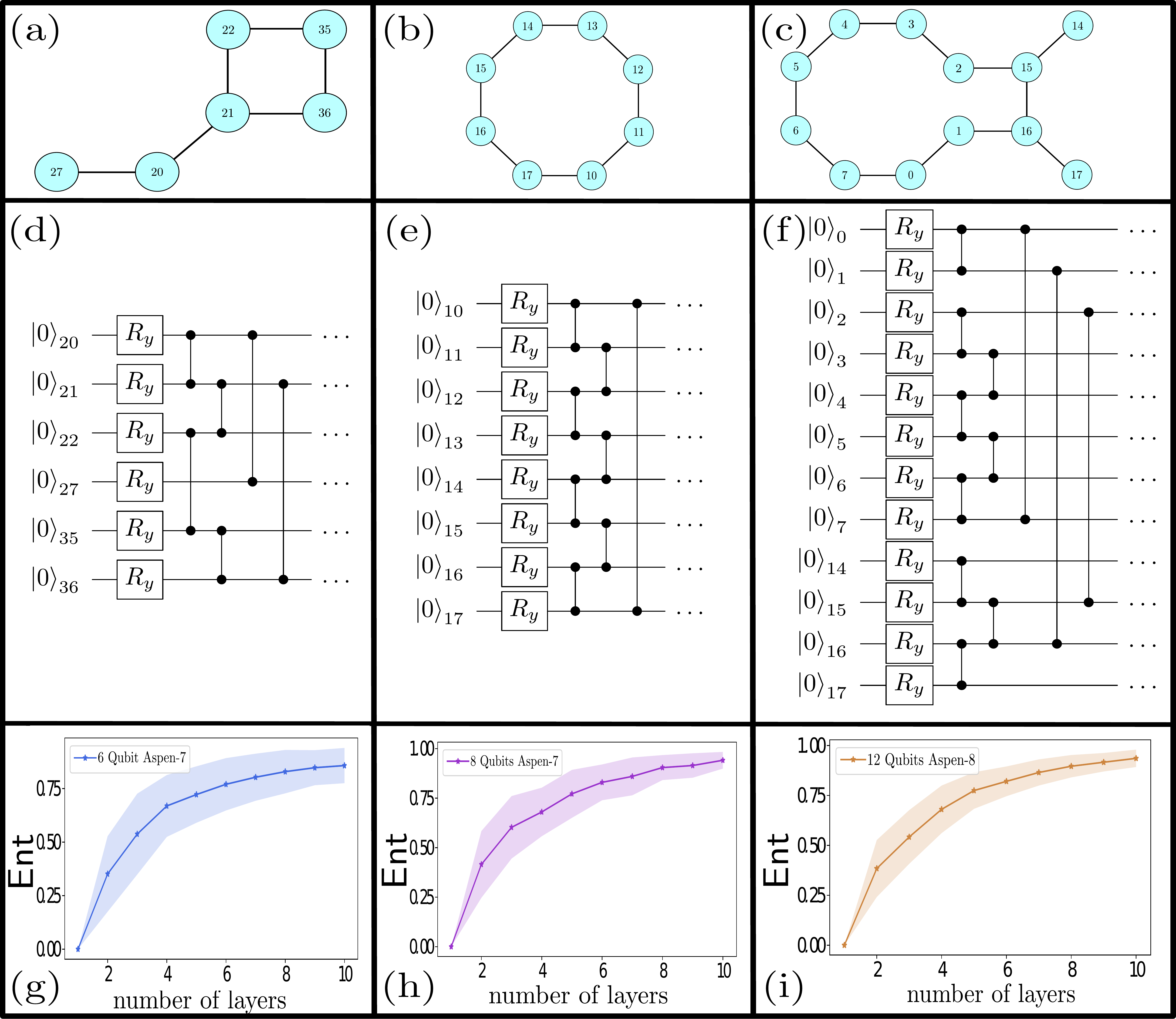}
    \caption{Hardware efficient circuits for $6$, $8$, $12$ qubit Born machine ansatz. (a)-(c) show \computerfont{Aspen-7-6Q-C},  \computerfont{Aspen-7-8Q-C} from the \computerfont{Aspen-7} chip and a $12$ qubit sublattice from the \computerfont{Aspen-8} chip which we consider. (d) - (f) illustrate the the entanglement structure in a single layer, which tightly matches the chip topology. (g) - (i)  show the average entangling capability, $\mathsf{Ent}$ in \eqref{eqn:average_ent_capability_definition} as a function of the number of layers in the circuit for each of the entangling structures shown in (d)-(f). Error bars show mean and standard deviation over $100$ random parameter instances, $\{\paramtheta_i\}_{i=1}^{100}, \paramtheta^j_i\sim U(0, 2\pi)$, in the single qubit rotations. $U$ is the uniform distribution over the interval $[0, 2\pi]$.}
    \label{fig:aspen_7_6q_8q_12q_circuit_ansatze}
\end{figure}

For each of the lattices in \Fref{fig:aspen_sublattices}, we fit the native entanglement structure using $\mathsf{CZ}$ gates, and layers of single qubits rotations. For convenience, we use $R_y$ rotation gates as the single qubit gates, which have the decomposition $R_{y}(\theta) = R_x(\pi/2)R_z(\theta)R_x(-\pi/2)$, using the Rigetti native single qubit rotations. The first `layer' contains only $R_y$ gates, and each layer thereafter consists of the hardware native $\mathsf{CZ}$ gates, plus a layer of $R_y$ gates. For the 4 qubit chip, \computerfont{Aspen-7-4Q-C}, we illustrate this in \Fref{fig:aspen_7_4q_circuit_ansatz}. For the other sublattices in \Fref{fig:aspen_sublattices}, we illustrate the entanglement structure in \Fref{fig:aspen_7_6q_8q_12q_circuit_ansatze} for the first layer of the circuits. In this way, an $n$ qubit QCBM with $l$ layers will have $n\times l$ trainable parameters.

For the above circuits, we compute the average Meyer-Wallach~\cite{meyer_global_2002} entanglement capacity, a measure of entanglement in quantum states proposed as a method of comparing different circuit $\Ansatze$ by \cite{sim_expressibility_2019}. This measure has been used in a similar context by \cite{hubregtsen_evaluation_2020} in order to draw connections between Ansatz structure and classification accuracy. The entanglement measure $Q$ is defined, for a given input state $\ket{\psi}$ as:
\begin{align}\label{eqn:meyer_wallach_entanglement_measure}
    Q(\ket{\psi}) := \frac{4}{n}\sum_{j=1}^n D(\iota_j(0)\ket{\psi}, \iota_j(1)\ket{\psi})\\
    D(\ket{\boldsymbol{u}}, \ket{\boldsymbol{v}}) = \frac{1}{2} \sum\limits_{i, j}|u_iv_j - u_jv_i|^2
\end{align}
where $D$ is a particular distance between two quantum states, $\ket{\boldsymbol{u}} := \sum_iu_i\ket{i}, \ket{\boldsymbol{v}} := \sum_jv_j\ket{j}$. This distance can be understood as the square of the area of the parallelogram created by vectors $\ket{\boldsymbol{u}}$ and $\ket{\boldsymbol{v}}$. The notation $\iota_j(b)$ is a linear map which acts on computational basis states as follows:
\begin{equation} \label{eqn:iota_definition}
    \iota_j(b)\ket{b_1\dots b_n} := \delta_{bb_j}\ket{b_1\dots \hat{b}_j\dots b_n}
\end{equation}
where $\hat{\cdot}$ indicates the absence of the $j^{th}$ qubit. For example, $    \iota_2(0)\ket{1001} = \ket{101}$
However, to evaluate $Q$ for a quantum state, we instead use the equivalent formulation derived by \cite{brennen_observable_2003}, which involves computing the purities of each subsystem of the state $\ket{\psi}$:
\begin{equation}\label{eqn:meyer_wallach_entanglement_alternative}
    Q(\ket{\psi}) = 2\left(1-\frac{1}{n}\sum_{k=1}^{n}\Tr[\rho_k^2]\right)
\end{equation}
where $\rho_k := \Tr_{\Bar{k}}\left(\ketbra{\psi}{\psi}\right)$ is the partial trace over every one of the $n$ subsystem of $\ket{\psi}$ \emph{except} $k$. This reformulation of $Q$ gives more efficient computation and operational meaning since the purity of a quantum state is efficiently computable. Given $Q$, we define \cite{sim_expressibility_2019} $\mathsf{Ent}$ as the average value of $Q$ over a set, $\mathcal{S}$ of $M$ randomly chosen parameter instances,  $S := \{\paramtheta_i\}_{i=1}^M$:
\begin{equation} \label{eqn:average_ent_capability_definition}
    \mathsf{Ent} := \frac{1}{|S|} \sum_{i} Q(\ket{\psi_{\paramtheta_i}})
\end{equation}
For the circuit $\Ansatze$ we choose, the value of $\mathsf{Ent}$ is plotted for a given number of layers in \Fref{fig:aspen_7_6q_8q_12q_circuit_ansatze}.

\subsection{Boltzmann Machine Structure} \label{ssec:boltzmann_machine_structure}

Given the above choice for a Born machine ansatz, we can build a corresponding restricted Boltzmann machine which has $n_{\mathrm{v}} := |\mathcal{N}_{\mathrm{v}}| = n$ visible nodes (where $n$ is the number of qubits) and $n_{\mathrm{h}} := |\mathcal{N}_{\mathrm{h}}| = nl-n = n\times (l-1)$ hidden nodes. To reiterate, we fix the RBM weights to have random values and only the local biases are trained. We revisit weight training in \ref{app_b_subapp:boltzmann_weight_training}.

\begin{figure}[t]
    \centering
    \includegraphics[width=0.8\columnwidth, height=0.3\columnwidth]{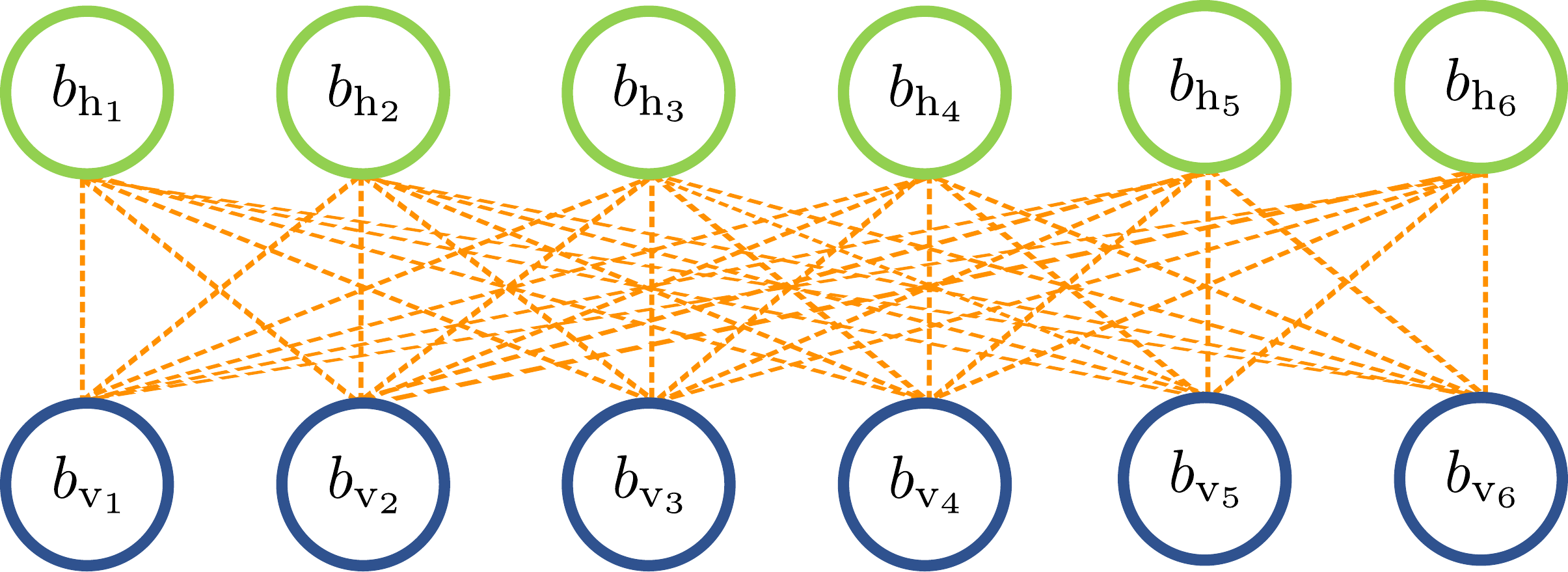}
    \caption{Restricted Boltzmann machine structure using $6$ visible and hidden nodes with $12$ parameters. Corresponds to the $6$ qubit Born machine in \Fref{fig:aspen_7_6q_8q_12q_circuit_ansatze}(d). Dotted lines indicate weights are not trainable but randomly chosen and fixed through training. Biases for visible and hidden nodes, $b_{\mathrm{v}_i}, b_{\mathrm{h}_i}$ correspond to the self-loop weights in \Fref{fig:gbm_architecture_examples} which are trainable.}
    \label{fig:rbm_4_node_structure}
\end{figure}

\section{Training Procedures} \label{sec:training_procedures}

In order to fit the model distribution to the data, one need some means of comparing how close these two distributions are. Typically, this comes in the form of a \emph{cost function}, $D(p_{\paramtheta}, \pi)$. In this work, we consider a variety of cost functions with which to compare both models we investigate. 

This cost function is then minimized during the training procedure to find a setting of the parameters, $\paramtheta$ such that $D(p_{\paramtheta}, \pi)$ is as small as possible. Gradient descent (GD) is a common method to minimize such costs in machine learning as it finds the steepest direction of descent in the parameter landscape defined by, $\paramtheta$. GD proceeds with a number of `epochs', where in each epoch ($t$) the parameters are updated as follows:
\begin{equation}\label{eqn:parameter_update_rule}
    \paramtheta^{(t+1)} \leftarrow \paramtheta^{(t)} - \Delta D(p_{\paramtheta^{(t)}}, \pi)
\end{equation}
$\Delta D(p_{\paramtheta}, \pi)$ is the update rule defining how each parameter should be updated, depending on the current value of $D$ and is negative since we wish to go downhill in the parameter landscape. The `vanilla' form of gradient descent simply directly uses an update of the form $\Delta D(p_{\paramtheta}, \pi) = \eta \partial_{\paramtheta^{(t)}} D(p_{\paramtheta}, \pi)$, where $\eta$ is a \emph{learning rate} and $\partial_{\paramtheta^{(t)}}D$ is the partial derivative of $D$ with respect to the current parameters. Computing this gradient efficiently can be a non-trivial procedure, and it is estimate given the data. More complicated update rules such as Adam \cite{kingma_adam_2015} are also possible, which include terms like `momentum' to the update rule, to improve convergence speed.

\subsection{Born Machine Training} \label{ssec:born_machine training}

The primary cost function we choose to train the Born machine is the Sinkhorn divergence (SHD), a recently defined~\cite{ramdas_wasserstein_2015, genevay_learning_2018, feydy_interpolating_2019} method of distribution comparison, and related to \emph{optimal transport} (OT)~\cite{villani_optimal_2009}, which is known to be a relatively powerful metric between probability distributions.
\begin{multline}
 \label{eqn:sinkhorndivergence}
      D(p_{\paramtheta}, \pi) := \mathcal{L}_{\mathrm{SHD}}^\epsilon(p_{\paramtheta}, \pi) \coloneqq\\
    \mathrm{OT}^c_\epsilon(p_{\paramtheta}, \pi) - \frac{1}{2} \mathrm{OT}^c_\epsilon(p_{\paramtheta}, p_{\paramtheta}) -\frac{1}{2}\mathrm{OT}^c_\epsilon(\pi, \pi) 
\end{multline}
\begin{multline}\label{eqn:wassersteinregularised}
    \mathrm{OT}^c_\epsilon(p_{\paramtheta}, \pi) \coloneqq 
      \min\limits_{U \in \mathcal{U}(p_{\paramtheta}, \pi)}\\
      \left(\sum\limits_{\substack{(\x, \y) \\ \in \mathcal{X}^d\times\mathcal{Y}^d}} c(\x, \y)U(\x, \y)  + \epsilon \mathrm{KL}(U|p_{\paramtheta}\otimes \pi)\right) 
\end{multline}
\noindent where $\epsilon \geq 0 $ is a regularisation parameter, and $\mathcal{U}(p_{\paramtheta}, \pi)$ is the set of all \textit{couplings} between $p_{\paramtheta}$ and $\pi$, i.e.\@ the set of all joint distributions, whose marginals with respect to $\x, \y$ are $p_{\paramtheta}(\x), \pi(\y)$ respectively. $\mathrm{KL}(U|p_{\paramtheta}\otimes \pi)$ is the Kullback-Leibler~\cite{kullback_information_1951} divergence (also relative entropy) between the coupling, $U$, and a product distribution composed of the model and the data, $p_{\paramtheta}\otimes \pi$. The introduction of the entropy term smooths the problem, so that it becomes more easily solvable, as a function of $\epsilon$.

We use this cost function since we numerically found it to be the best choice, in terms of speed and accuracy of training. However, we provide a comparison to the \emph{maximum mean discrepancy} ($\mathrm{MMD}$) cost function, training with respect to an adversarial discriminator and a gradient free genetic algorithm in \ref{app_a:alternative_born_training}.

As shown in ~\cite{coyle_born_2020} we can derive gradients of the Sinkhorn divergence, with respect to the given parameter, ${\paramtheta}_k$, since each parameterised gate we employ has the form $U(\theta)=\exp(\mathrm{i}\theta/2\Sigma)$, where $\Sigma^2 = \mathds{1}$. Using the parameter shift rule~\cite{mitarai_quantum_2018, schuld_evaluating_2019}, the gradient can be written as follows:
\begin{align}  \label{eqn:sinkhorn_gradient}
    \frac{\partial \mathcal{L}_{\textrm{SHD}}^\epsilon(p_{\paramtheta}, \pi)}{\partial {\paramtheta}_k} &= \sum\limits_{\x} \frac{\partial \mathcal{L}_{\textrm{SHD}}^\epsilon(p_{\paramtheta}, \pi)}{\partial p_{\paramtheta}(\x)}\frac{\partial p_{\paramtheta}(\x)}{\partial {\paramtheta}_k} \\
    &= \frac{1}{2}\sum\limits_{\x}\varphi(\x)\left(p_{{\paramtheta}^+_k}(\x) - p_{{\paramtheta}^-_k}(\x) \right)  \\
    & = \frac{1}{2}\left(\underset{\substack{\x \sim p_{{\paramtheta}_k^+}  }}{\mathbb{E}}[\varphi(\x)] - \underset{\substack{\x \sim p_{{\paramtheta}^-} }}{\mathbb{E}}[\varphi(\x)] \right)
\end{align}
The function $\varphi$ is defined\cite{feydy_interpolating_2019} in order to ensure the gradient extends to the entire sample space, and is defined as follows for each sample, $\x$:
\begin{multline} \label{eqn:sinkhorn_function_definition}
    \varphi(\x) =\\
    -\epsilon \text{LSE}_{k=1}^M\left(\log\left(\pi(\mathbf{\y}^k) + \frac{1}{\epsilon}g(\y^k) - \frac{1}{\epsilon} C(\x, \y^k)\right)\right)\\
    + \epsilon \text{LSE}_{k=1}^N\left(\log\left(p_{\paramtheta}(\x^k) + \frac{1}{\epsilon}s(\x^k) - \frac{1}{\epsilon} C(\x, \x^k)\right)\right)
\end{multline}
Therefore, one can compute the gradient by drawing samples from the distributions, $\hat{\x} \sim p_{{\paramtheta}^\pm}$, and computing the vector $\varphi(\x)$, for each sample, $\x$. The functions $g$ and $s$ in \Eref{eqn:sinkhorn_function_definition} are optimal Sinkhorn potentials, arising from a primal-dual formulation of optimal transport. These are computed using the Sinkhorn algorithm, which gives the divergence its name~\cite{sinkhorn_relationship_1964}. $C(\x, \y)$ is the optimal transport \textit{cost matrix} derived from the cost function applied to all samples, $C_{ij}(\x^i, \y^j) = c(\x^i, \y^j)$ and $\text{LSE}_{k=1}^N(\boldsymbol{V}_k) = \log\sum\limits_{k=1}^N\exp(\boldsymbol{V}_k)$ is a log-sum-exp reduction for a vector $\boldsymbol{V}$. For further details on how the functions, $g$ and $s$ are computed, see along with the Sinkhorn divergence and its gradient, see~\cite{feydy_interpolating_2019, coyle_born_2020}.

\subsection{Boltzmann Machine Training} \label{ssec:boltzmann_training}

For the RBM, we use the standard Boltzmann protocol of maximizing the log-likelihood function $\mathcal{L}$
\footnote{Equivalent to minimizing an empirical cost $\widetilde{D}(p_{\paramtheta}, \pi) = 1 - \mathcal{L}(Y, \paramtheta)$. The maximization procedure adds an extra negative sign to the update rule.}, i.e. the probability of generating vectors belonging to a training set $Y = \{\boldsymbol{y}\}$:
\begin{equation}
\mathcal{L}(Y, \paramtheta) = \log(p_{\paramtheta}(Y))
\end{equation}
where $\paramtheta$ are the Boltzmann machine model parameters and $p(\boldsymbol{y})$ is the probability of generating data vectors $\boldsymbol{y}\sim \pi(\boldsymbol{y})$. For a particular data vector $\boldsymbol{y} \in Y$, we can take the likelihood function $\mathcal{L}$ as our cost function as a function of the model parameters $\paramtheta = \mathcal{W} =\{W_e\}$ \cite{fischer_training_2014}, which results in the gradient: 
\begin{equation}  \label{eqn:boltz_update_equation}
\frac{\partial \mathcal{L}}{\partial \paramtheta} = \frac{\partial \mathcal{L}}{\partial W_{e}} = \langle v_{e} \rangle_{\pi} - \langle v_{e} \rangle_{p_{\paramtheta}}
\end{equation}
wherein $W_{e}$ of \eqref{eqn:gbm_energy} are the model parameters coupling their respective set of nodes $v_{e} := \prod_{v_{i} \in e} v_{i}$, $\langle v_{e} \rangle_{\pi}$ and $\langle v_{e} \rangle_{p_{\paramtheta}}$ are the expectation values of $v_{e}$ calculated from the data and model distributions respectively where $v_{e}$ is taken to be a random variable taking values $\{0, 1\}$. 

As an example, consider an update to a (visible node) bias term (i.e.\@ a self-loop in \Fref{fig:gbm_architecture_examples}), we have $W_e = b_{\mathrm{v}_i}$ and also $v_{e}$ is simply $b_{\mathrm{v}_i}$ since the edge connects only one node. Then the gradient is computed using the expectation value of the bias:
\begin{equation}  \label{eqn:boltz_update_equation_bias_example}
\frac{\partial \mathcal{L}}{\partial b_{\mathrm{v}_i}} = \langle b_{\mathrm{v}_i} \rangle_{\pi} - \langle b_{\mathrm{v}_i} \rangle_{p_{\paramtheta}}
\end{equation}

In this work, we use vanilla gradient descent as the update rule, but we note we also considered more complex update rules or optimizers such as Adam~\cite{kingma_adam_2015}, and we found that this only improved the convergence speed, and not the final accuracy of the model.

The above is discussion has no quantum component, as the update rule and the model is completely classical. However, in order to actually compute the first and second order moment terms (using $M$ data and $N$ model-generated bitstrings) in \eqref{eqn:boltz_update_equation}, we require a method of generating samples from the RBM.
Unlike the Born machine, sample generation from a Boltzmann machine is not a trivial matter. Typical approaches are based on Gibbs sampling, for example $k$-step contrastive divergence \cite{hinton_practical_2012}. Here,  we use a method called QxSQA, a GPGPU-Accelerated Simulated quantum annealer based on Path-Integral Monte Carlo (PIMC) \cite{padilha_qxsqa_2019}. This simulated QA has been shown to be useful for sampling Boltzmann-like distributions, and we have shown the ability to use this sampling to train large \emph{quantum} generalized Boltzmann machines (QGBMs) for the purposes of generating synthetic data based on images \cite{lecun_gradient-based_1998} and financial data \cite{kondratyev_market_2019, kondratyev_non-differentiable_2020} in previous research \cite{henderson_generation_2019}.

\section{Results} \label{sec:results}

Here we present the numerical results obtained above using the models and training methods detailed above. In particular, we focus of training using the Sinkhorn divergence with the Adam optimiser~\cite{kingma_adam_2015} and its analytic gradients for the Born Machine, and log-likelihood maximization using QxSQA for training the Boltzmann machine. In \ref{app_a:alternative_born_training}, we revisit alternative training methods. As mentioned above, in the first instance, we also fix both models to only have trainable local parameters for simplicity. For the Born machine, this corresponds to only training single qubit unitaries, with the two qubit gates being unparameterised, and for the Boltzmann machine, this corresponds to training the biases of each node. We force each model to have the same number of trainable parameters in this way. The entanglement structure in the Born machine is fixed by the problem size, via the lattice topology, and the weights of the RBM are chosen to be random (but fixed) values on each instance. It is difficult to directly compare the connectivity of the models, howeveer we also experimented with randomly pruning the RBM weights to enforce the same number of connections as in the QCBM, but we found this did not affect performance significantly.

As a method of benchmarking the expressive power of each model in a fair way, we use an adversarial discriminator, and judge the performance relative to it. Specifically, we use a random forest discriminator from \computerfont{scikit-learn} \cite{pedregosa_scikit-learn_2011} with $1000$ estimators. A higher discriminator error implies better performance, with an error of $50\%$ indicates the discriminator can at best guess randomly when presented with a sample as to its origin - whether it came from the real data, or the model. Where error bars are shown, they correspond to the mean and standard deviations of the training over 5 independent runs. As the QCBM scales, the classical simulation becomes a bottleneck and limits the number of runs which can be done.

\begin{figure}[t]
    \centering
    \includegraphics[width=\columnwidth, height=0.8\columnwidth]{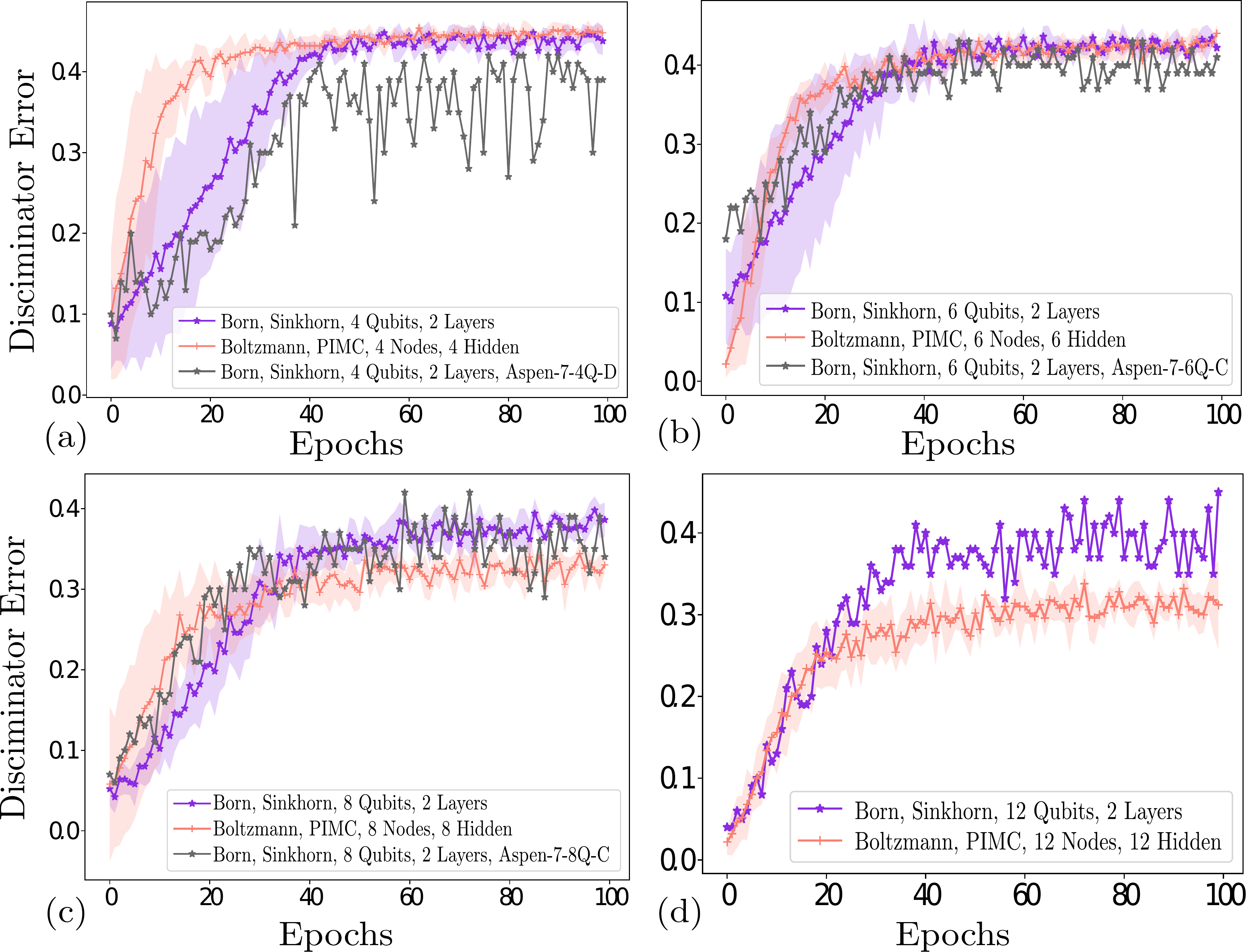}
    \caption{2 currency pairs (specifically \textsf{EUR/USD} and \textsf{GBP/USD}) at (a) 2 bits, (b) 3 bits, (c) 4 bits and (d) 6 bits of precision. Correspondingly, we use a QCBM of $4, 6, 8$ and $12$ qubits using the Ans\"{a}tze described above, and an RBM with the same numbers of visible nodes. The hidden units are scaled in each case to match 2 layers of the QCBM. Results when the QCBM is run on sublattices of the \computerfont{Aspen-7} QPU are shown in grey.}
    \label{fig:2_currency_pairs_born_v_boltz}
\end{figure}

\begin{figure}[t]
    \centering
    \includegraphics[width=\columnwidth, height=0.4\columnwidth]{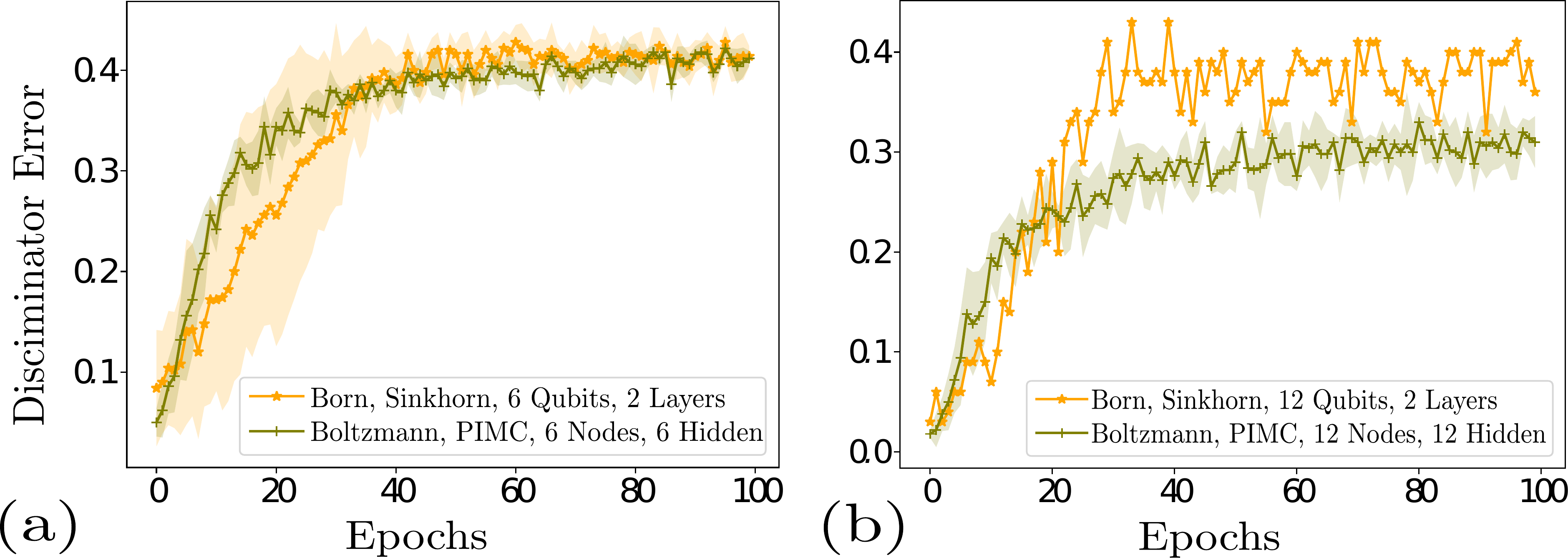}
    \caption{3 currency pairs at (a) 2 bits and (b) 4 bits of precision, using a QCBM of $6$ and $12$ qubits and an RBM with the same numbers of visible nodes.}
    \label{fig:3_currency_pairs_born_v_boltz}
\end{figure}

\begin{figure}[t]
    \centering
    \includegraphics[width=\columnwidth, height=0.4\columnwidth]{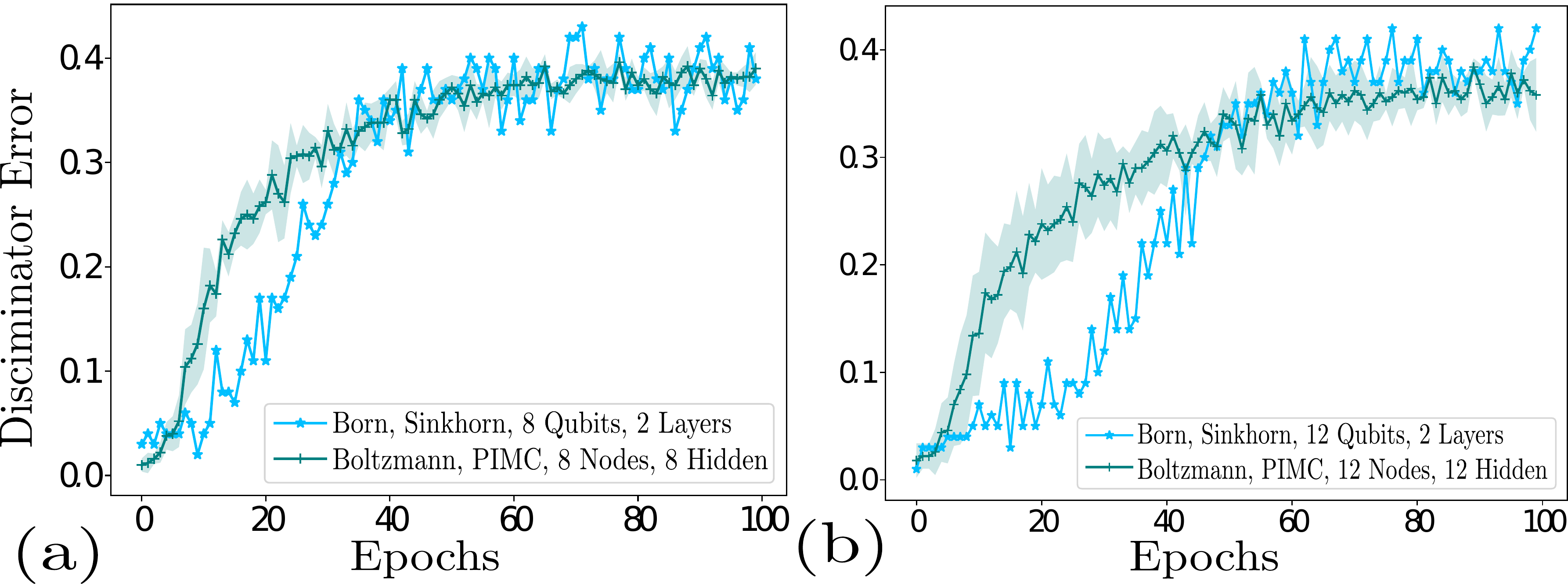}
    \caption{All $4$ currency pairs at (a) 2 bits, (b) 3 bits of precision, using a QCBM of $8$ and $12$ qubits and RBM with the same numbers of visible nodes. We notice the RBM again performs similarly to the Born machine for 2 bits of precision, but begins to be outstripped by it at 3 bits.}
    \label{fig:4_currency_pairs_born_v_boltz}
\end{figure}

\begin{figure}[t]
    \centering
    \includegraphics[width=\columnwidth, height=0.5\columnwidth]{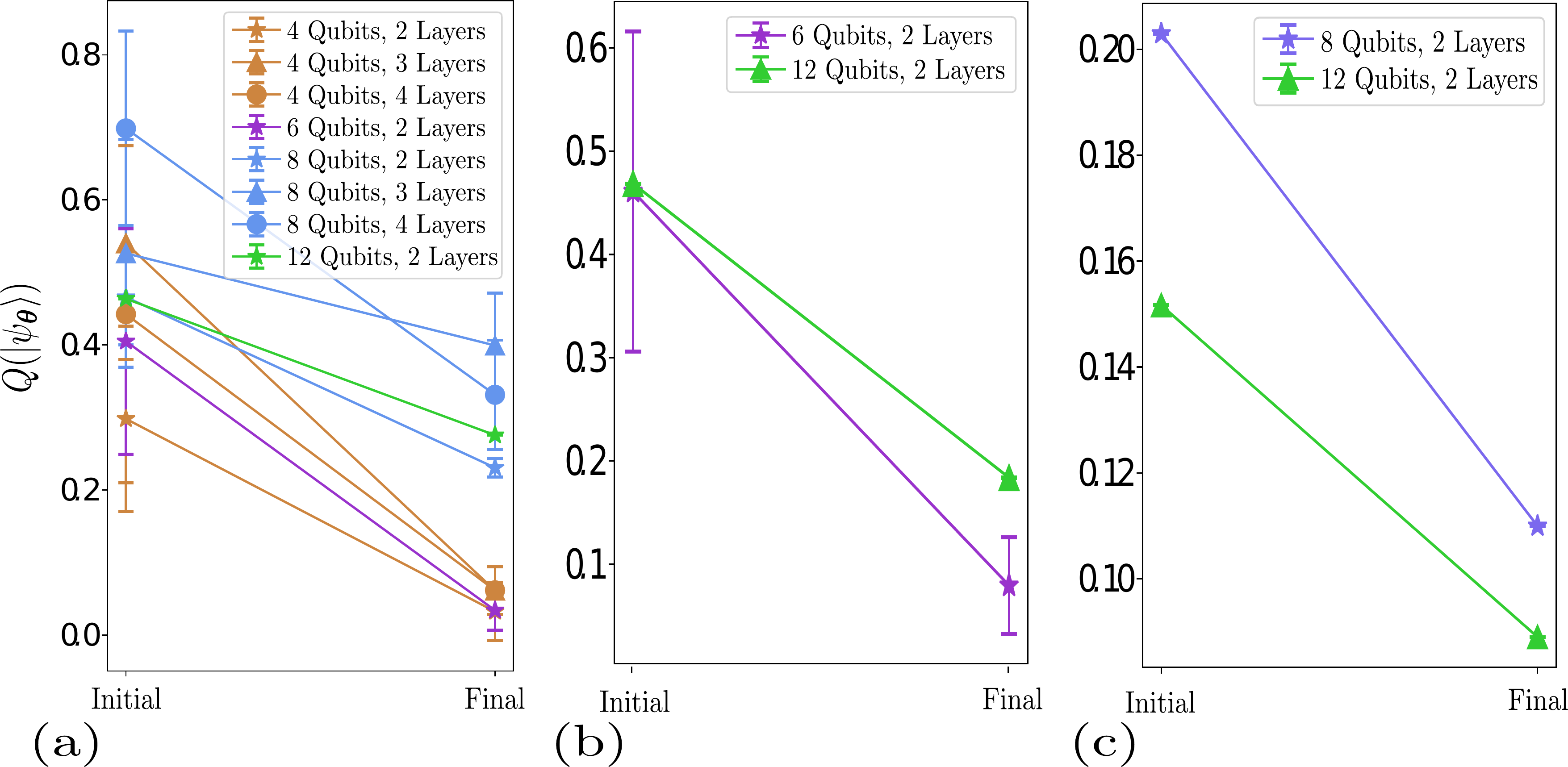}
    \caption{Meyer-Wallach entangling capability \eref{eqn:meyer_wallach_entanglement_alternative} for a random choice of parameters (Initial) and the trained parameters (Final) in the same circuit. Error bars represent mean and standard deviation over 5 independent training runs, where they are shown. The circuit $\ansatze$ used are those above in \Fref{fig:aspen_7_6q_8q_12q_circuit_ansatze} closely matching the corresponding chip topology. In each panel we see the circuits trained on (a) $2$ currency pairs at $2, 3, 4, 6$ bits of precision, (b) $3$ currency pairs at $2$ and $4$ bits of precision and (c) $4$ currency pairs at $2$ and $3$ bits of precision.}
    \label{fig:entangling_capability_in_training}
\end{figure}

In summary, we find the Born machine has the capacity to outperform the RBM as the precision of the currency pairs increases. In \Fref{fig:2_currency_pairs_born_v_boltz}, we use data from $2$ currency pairs, at $2, 3, 4$ and $6$ bits of precision. We notice the Born machine outperforms the RBM around $4$ bits (measured by a higher discriminator error), and still performs relatively well when run on the QPU. Similar bahaviour is observed for $3$ currency pairs in \Fref{fig:3_currency_pairs_born_v_boltz}, which uses a precision of $2$ and $4$ bits, and with $4$ pairs in \Fref{fig:4_currency_pairs_born_v_boltz} for a precision of $2$ and $3$ bits. In \Fref{fig:entangling_capability_in_training} we plot the entangling capability (defined by \eref{eqn:meyer_wallach_entanglement_alternative}) of the states generated by initial and final circuits learned via training. Curiously, we notice that in the problem instances in which the Born machine outperforms the Boltzmann machine (those with a higher level of precision), the trained circuits have a higher level of entanglement than those that do not, despite the data being completely classical in nature. This is especially prominent for $2$ currency pairs in \Fref{fig:entangling_capability_in_training}(a), in which the training drives the entanglement capability at $2$ and $3$ bits of precision close to zero (even for increased numbers of layers), but it is significantly higher for $4$ and $6$ bits of precision, when the Born machine outperforms the RBM, as seen in \Fref{fig:2_currency_pairs_born_v_boltz}. Similar behaviour is seen for $3$ currency pairs, but not as evident for $4$ pairs. The latter effect is possibly correlated to the similar performance of both models for $4$ currency pairs up to $3$ bits of precision.

\begin{figure}[t]
    \centering
    \includegraphics[width=\columnwidth, height=0.45\columnwidth]{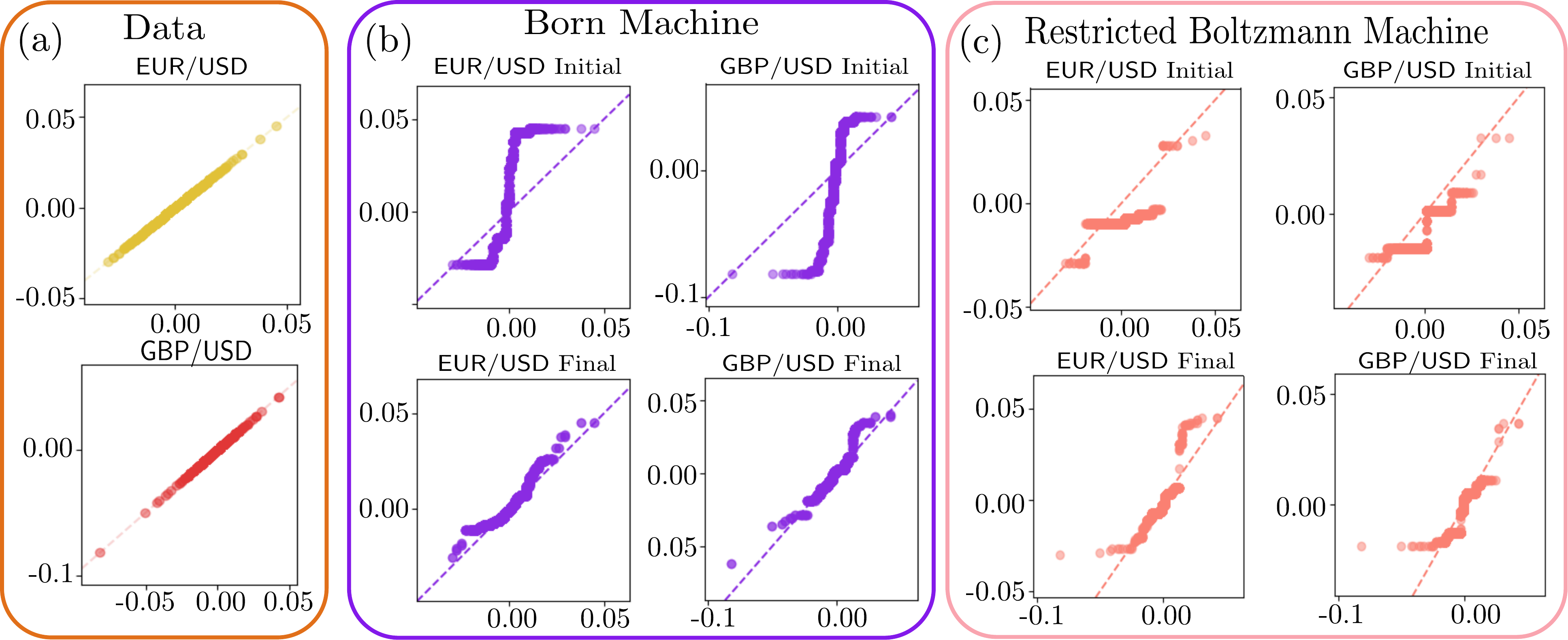}
    \caption{QQ Plots corresponding to \Fref{fig:2_currency_pairs_born_v_boltz}(d) of the marginal distributions of $2$ currency pairs ($\mathsf{EUR/USD}$ and $\mathsf{GBP/USD}$) at $6$ bits of precision. The Born machine distributions (purple) and those generated by the Boltzmann machine in (pink). (a) shows the QQ plot for the marginal distribution of each currency pair with respect to itself as a benchmark. (b) Born machine initial (top panels) and final (bottom panels) marginal distributions for both pairs and similarly in (c) for the RBM. While not able to completely mimic the data due to the low number of parameters, the Born machine clearly produces a better fit.}
    \label{fig:QQ_plots_2_pairs}
\end{figure}

\begin{figure}[t]
    \centering
    \includegraphics[width=\columnwidth, height=0.6\columnwidth]{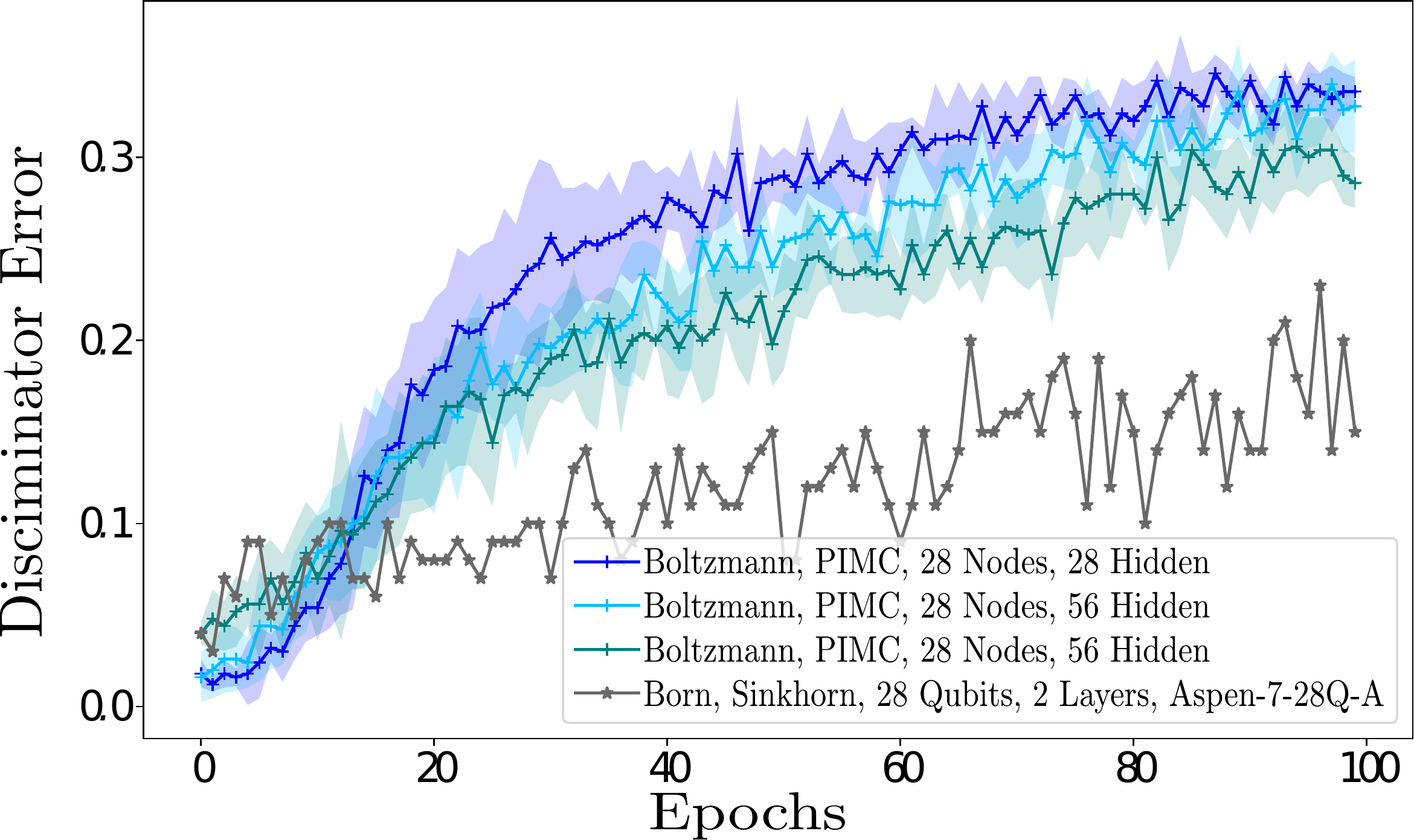}
    \caption{Random forest discriminator during training for a problem size of $4$ currency pairs at $7$ bits of precision, using $28$ visible nodes in the Boltzmann machine and $28$ qubits in the Born machine. The $28$ qubit Born machine is run exclusively on the \computerfont{Aspen-7-28Q-A} chip using 2 layers of the hardware efficient ansatz similar to those shown in \Fref{fig:aspen_7_6q_8q_12q_circuit_ansatze}.}
    \label{fig:28_qubit_born_boltzmann}
\end{figure}

We are also able to somewhat successfully train the largest instance of a Born Machine
to date in the literature, namely one consisting of $28$ qubits on the Rigetti \computerfont{Aspen-7} chip using the Sinkhorn divergence (whose topology is shown in \Fref{fig:aspen_sublattices}(e), and we find it performs surprisingly well. We show the performance of the $28$ qubit model versus the a Boltzmann machine with $28$ visible nodes, and a suitable number of hidden nodes to match the number of parameters in the Born machine in \Fref{fig:28_qubit_born_boltzmann}. While the performance of the Born machine is significantly less than that of its counterpart, it is clear that the model is learning (despite hardware errors), up to a discriminator error of $20\%$. While this result seems to contradict the previous findings in this work, we emphasize that it does not, since we are not able to simulate the QCBM at this scale in a reasonable amount of time. We would not necessarily expect the Born machine to match the performance of the RBM \emph{on hardware} at this scale for a number of reasons, the most likely cause for diminishing performance is quantum errors in the hardware. However we cannot rule out other factors, such as the Ansatz choice. We leave thorough investigation of improving hardware performance to future work, perhaps by including error mitigation~\cite{hamilton_error-mitigated_2019} to reduce errors, parametric compilation and active qubit reset~\cite{smith_practical_2016, karalekas_quantum-classical_2020} to improve running time and other techniques.

\section{Discussion}\label{sec:discussion}
In conclusion, we investigate and compare two different models when trained on a real-world financial dataset consisting of currency pairs at varying levels of precision. We chose a completely classical model in the restricted Boltzmann machine, and put it up against a completely quantum model in the form of a quantum circuit Born machine, in order to compare their relevant expressive powers, and supplement recent related work in this direction~\cite{kondratyev_non-differentiable_2020, alcazar_classical_2020}. As a benchmark of fairness, we fixed the models to have the same numbers of trainable parameters and found that the simulated Born machine always performed at least as well as the RBM, and in several cases outperformed it, measured relative to the accuracy of an adversarial discriminator. To complement this finding, we investigated the entangling capability of the circuits learned by the QCBM, and found a rough correlation between training towards higher levels of entanglement, and outperforming the classical model.

From this work, there are many possible avenues for exploration. The first, is improving the Born machine training speed by, for example, leveraging GPU accelerated computation of the cost functions, and also incorporating techniques to improve running time and execution on the QPU. Furthermore, to improve performance, one could consider variable structure $\Ansatze$ \cite{cincio_learning_2018, cerezo_variational_2020} or quantum-specific optimizers~ \cite{kubler_adaptive_2020, arrasmith_operator_2020, lavrijsen_classical_2020} for the model and training. An alternative direction, is to enlarge the suite of classical model comparison to compare the Born machine to, in order to solidify any perceived advantage and extending the model into mixed states to potentially increase the expressive power~\cite{verdon_quantum_2019}. Alternatively, one could investigate methods to divide the classical-quantum resources in the learning procedure~\cite{paini_approximate_2019}.

\ack
We thank Alexei Kondratyev for useful discussions and comments, as well as the Rigetti Computing Aspen-7 hardware team for access necessary for the experiments in this research. This work was supported by the Engineering and Physical Sciences Research Council (grant EP/L01503X/1), EPSRC Centre for Doctoral Training in Pervasive Parallelism at the University of Edinburgh, School of Informatics and Entrapping Machines, (grant FA9550-17-1-0055).

\bibliographystyle{unsrt}

\appendix

\section{Alternative Training Methods} \label{app_a:alternative_born_training}

Here we provide numerical results illustrating the training of the Born machine using some alternative methods and cost functions, for small numbers of qubits.

\subsection{Maximum Mean Discrepancy} \label{ssec:mmd}

The first alternative method is derived by using a different cost function, the so-called maximum mean discrepancy ($\mathrm{MMD}$). Like optimal transport, this defines a metric on the space of probability distributions, and from which, an efficient-to-compute method of comparison can be defined~\cite{borgwardt_integrating_2006, gretton_kernel_2007}:
\begin{multline}
  D(p_{\paramtheta}, \pi) \coloneqq \mathcal{L}_{\mathrm{MMD}}  \coloneqq \\ \underset{\substack{\mathbf{x} \sim  p_{\boldsymbol\theta} \\\mathbf{y} \sim p_{\boldsymbol\theta}}}{\mathbb{E}}(\kappa(\mathbf{x},\mathbf{y})) + \underset{\substack{\mathbf{x} \sim \pi \\\mathbf{y} \sim \pi }}{\mathbb{E}}(\kappa(\mathbf{x},\mathbf{y})) -\underset{\substack{\mathbf{x} \sim p_{\boldsymbol\theta}\\ \mathbf{y} \sim \pi}}{2\mathbb{E}}(\kappa(\mathbf{x},\mathbf{y})) \label{mmdexact}
\end{multline}
This cost function was originally utilized for hypothesis testing~\cite{gretton_kernel_2007}, but has since found use in training generative models. In particular, it enabled the first approach to train a QCBM~\cite{liu_differentiable_2018, coyle_born_2020} in a differentiable way. 

The function $\kappa$ is a \emph{kernel}, which enables a means of comparison on the support spaces, $\mathcal{X}, \mathcal{Y}$. For this work, we choose the common Gaussian mixture kernel~\cite{liu_differentiable_2018} for the $\mathrm{MMD}$, which is universal, and hence enables the $\mathrm{MMD}$ to act as a faithful method of distribution comparison:
\begin{equation}
    \kappa_G(\mathbf{x}, \mathbf{y}) \coloneqq  \frac{1}{c}\sum_{i=1}^c\exp\left(-\frac{||\mathbf{x}-\mathbf{y}||^2_2}{2\sigma_i}\right) \label{gaussiankernel}
\end{equation}
The parameters, $\sigma_i$, are \emph{bandwidths} which determine the scale at which the samples are compared, and $||\cdot||_2$ is the $\ell_2$ norm. Here we choose $\sigma = [0.25, 10, 1000]$, as in \cite{liu_differentiable_2018}. Typically, the kernel is a classical function, but quantum kernels can also be considered here~\cite{coyle_born_2020, kubler_quantum_2019, schuld_quantum_2019, havlicek_supervised_2019}

\subsection{Adversarial Discriminator} \label{ssec:adversarial_discrim_training}

The second method we can choose to use is to not only use a discriminator as a benchmark, but also to train the model relative to it. As in the above cases, this is a gradient based approach, with the analytic gradient taken by differentiating the discriminator loss.

Adversarial training has become a popular and powerful way to train neural networks, originating with generative adversarial networks (GANs) \cite{goodfellow_generative_2014}. GANs are composed of two machine learning components, a discriminator, $\mathcal{D}$, which attempts to predict if a sample $\boldsymbol{x}$ is from a data distribution or rather has been generated by a generator network $\mathcal{G}$ (in our notation, the generator network samples from a probability distribution, $p_{\paramtheta}$ and is either a Born machine or a Boltzmann machine). Generalizations of the GAN into the quantum domain have also been considered~\cite{romero_variational_2019, lloyd_quantum_2018, dallaire-demers_quantum_2018, zoufal_quantum_2019, anand_experimental_2020}. The generator attempts to minimize the following loss:
\begin{equation} \label{eqn:adversarial_generator_cost}
\mathcal{L}_{\mathcal{G}} := \mathbb{E}\left[ \log \left(1 - \mathcal{D}\left(\x\right) \right)\right] \approx \frac{1}{N}\sum\limits_{i=1}^N\left[ \log \left(1 - \mathcal{D}\left(\x_i\right) \right)\right]
\end{equation}
where $\mathcal{D}\left(\boldsymbol{x}\right)$ is the probability that a discriminator, $\mathcal{D}$, guesses that $\x$ is from the real data set. The approximation to the expectation value is taken over $N$ generated samples in practice. In order to train the generator with respect to this cost function (taken with respect to a specific discriminator, $\mathcal{D}$), gradient descent can be used to minimize \eref{eqn:adversarial_generator_cost}, with the gradient given by:
\begin{multline} \label{eqn:adversarial_generator_update}
    \nabla_{\paramtheta} \mathbb{E}\left[\log \left(1-\mathcal{D}\left(\x_i\right)\right)\right] =\\ \nabla_{\paramtheta} \sum_{\x}p_{\paramtheta}(\x)\left[\log \left(1-\mathcal{D}\left(\x_i\right)\right)\right] =\\ \sum_{\x}\nabla_{\paramtheta} p_{\paramtheta}(\x)\left[\log \left(1-\mathcal{D}\left(\x_i\right)\right)\right]
\end{multline}
If we again assume the generator network is a Born machine, composed of quantum gates of the form $U(\theta)=\exp(\mathrm{i}\theta/2 \Sigma)$, then using the parameter shift rule as for the Sinkhorn divergence above \eref{eqn:sinkhorn_gradient}, we get:
\begin{multline}  \label{eqn:adversarial_generator_gradient}
\nabla_{\paramtheta} \mathcal{L}_{\mathcal{G}}  = \frac{1}{2}\sum_{\x}p_{\paramtheta^+}\left[\log \left(1-\mathcal{D}\left(\x_i\right)\right)\right] -\\ p_{\paramtheta^-}\left[\log \left(1-\mathcal{D}\left(\x_i\right)\right)\right]
\end{multline}
These expectation values can be evaluated by sampling from the parameter shifted circuit distributions, $p_{\paramtheta^{\pm}}$ as usual. Correspondingly, while a generator is trying to minimize the above cost, \eref{eqn:adversarial_generator_cost}, the discriminator can also be trained for a number of sub-steps to become better at identifying false samples. This can be done by using gradient \emph{ascent} to maximize the following cost:
\begin{equation}\label{eqn:adversarial_discriminator_cost}
    \mathcal{L}_{\mathcal{D}} \approx \frac{1}{M}\sum_{j=1}^{M}\log \mathcal{D}\left(\y_j\right) + \frac{1}{N}\sum_{i=1}^{N}\log \left(1- \mathcal{D}\left(\x_i\right)\right)
\end{equation}
where the latter term is the same as in \eref{eqn:adversarial_generator_update}, and the former represents the probability that $\mathcal{D}$ is able to correctly identify true data samples, $\y \sim \pi$. The gradient of \eref{eqn:adversarial_discriminator_cost} can be computed similarly.

In this work, we implement the training laid out in this section with two slight variations to note:
\begin{enumerate}
  \item We used a slightly revised version of Eq. \ref{eqn:adversarial_generator_gradient} which dropped the 0.5 and log components (simply used $1- \mathcal{D}$ in both terms). As we were still using Adam as the update optimizer, we believe that asserting this overall should not pose any major impact to performance. Moreover, the adversarial approach was still slower compared to the Sinkhorn divergence, and therefore did not garner increased focus in this work.
  \item As the modeling problem in this work was extremely small, we choose to simply re-train a new discriminator model from scratch at every generative model training iteration, with corresponding test set error of the discriminator being recorded and used as a primary metric in this work. Similarly, a different discriminator model was used for calculating model parameter updates every training iteration while using adversarial training.
\end{enumerate}

\begin{figure}[ht!]
    \centering
    \includegraphics[width=\columnwidth, height=0.35\columnwidth]{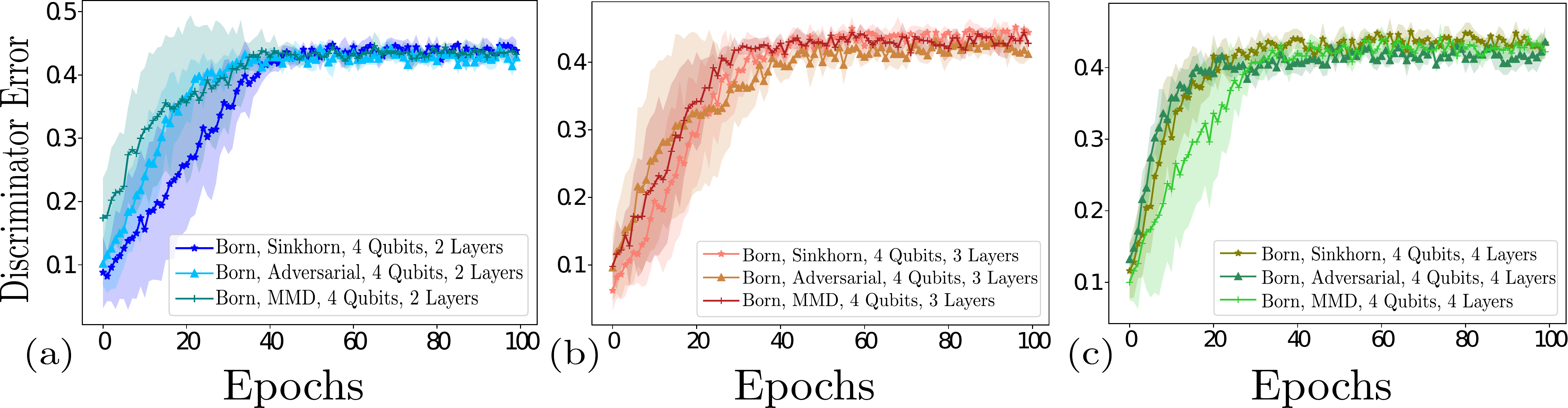}
    \caption{Training with the Sinkhorn divergence, the MMD, and the adversarial discriminator for (a) $2$, (b) $3$, (c) $4$ layers of the hardware efficient ansatz for $4$ qubits.}
    \label{fig:sinkhorn_v_mmd_v_adversarial_training}
\end{figure}

\subsection{Genetic Algorithm} \label{ssec:genetic_algorithm}

Finally, we use a gradient free approach in a genetic algorithm, since this was also used to train a QCBM on this same dataset~\cite{kondratyev_non-differentiable_2020}. One could also choose one of the many gradient free optimisers from \computerfont{scikit-learn}, as has been also done for Born machines~\cite{benedetti_generative_2019}. A simplified version of a genetic algorithm was implemented in ~\cite{kondratyev_non-differentiable_2020} due to the low number of parameters in a 12 qubit Born machine. We found that this method was significantly slower than the gradient based methods we discuss above.

\begin{figure}[ht!]
    \centering
    \includegraphics[width=\columnwidth, height=0.28\columnwidth]{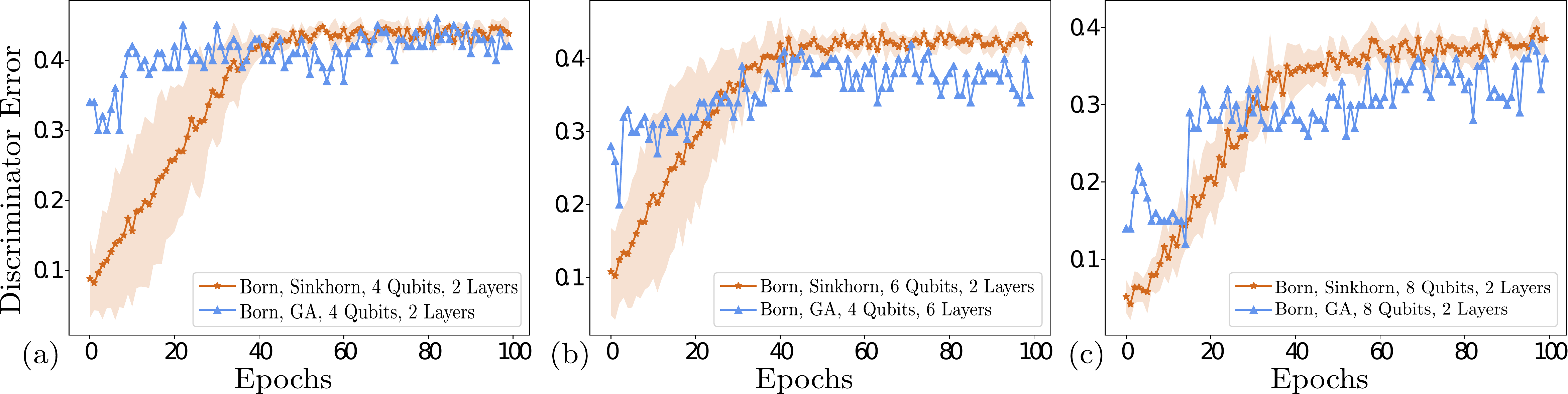}
    \caption{Training  a QCBM with $4, 6$ and $8$ qubits using the Sinkhorn divergence (gradient based) versus a genetic algorithm (gradient free).}
    \label{fig:sinkhorn_v_genetic_training}
\end{figure}

\section{Alternative Model Structures} \label{app_b:alternative_model_structure}

Here we showcase the effect of using alternative model structures for the QCBM and the RBM.

\subsection{Differing numbers of Born machine layers} \label{app_a_subapp:differing_number_layers}

\begin{figure}[ht!]
    \centering
    \includegraphics[width=\columnwidth, height=0.4\columnwidth]{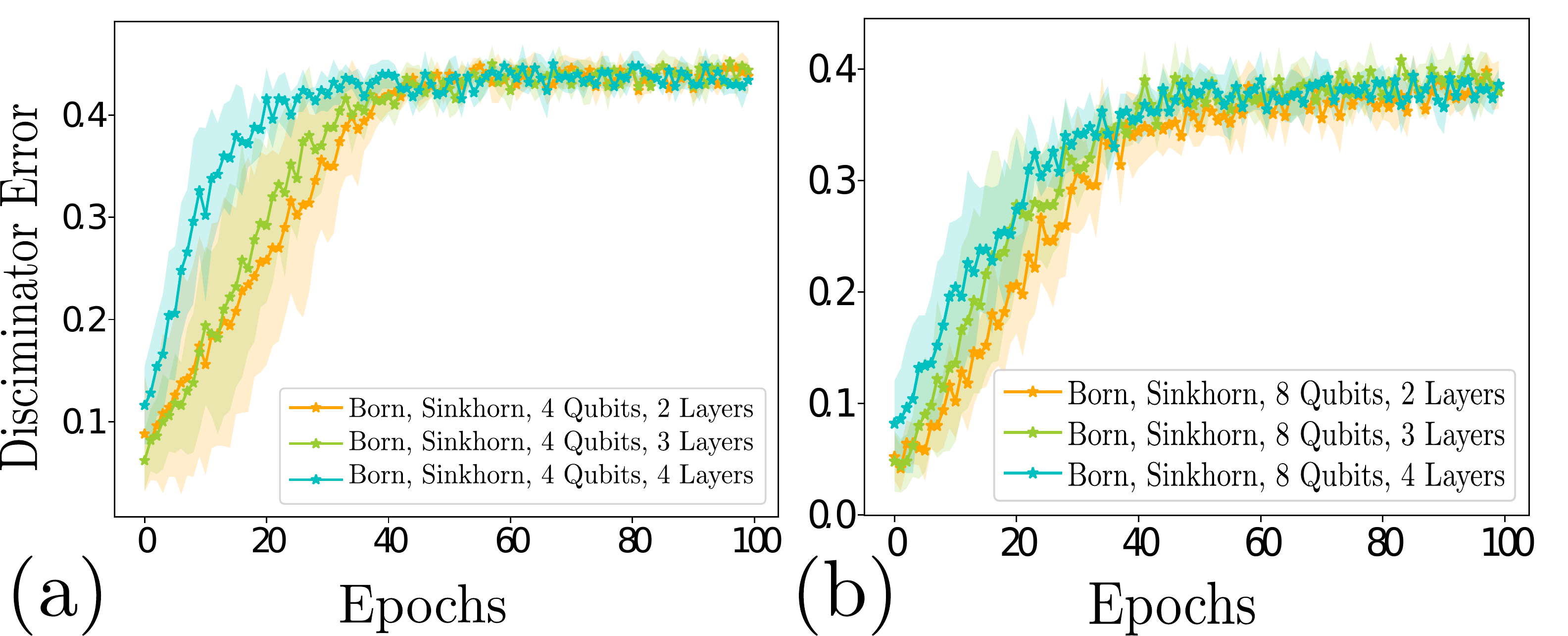}
    \caption{$2, 3$ and $4$ layers of the hardware efficient ansatz for (a) $4$ and (b) $8$ qubits. Models are trained on $2$ currency pairs at $2$ and $4$ bits of precision respectively. No major advantage observed for using an increasing number of layers, except perhaps in convergence speed, suggesting that $2$ layers is sufficient for these problem instances.}
    \label{fig:differing_layers}
\end{figure}

For completeness, in \Fref{fig:differing_layers}, we show the effect of alternating the number of layers of the hardware efficient $\Ansatze$, shown in \Fref{fig:aspen_7_6q_8q_12q_circuit_ansatze} for $4$ and $8$ qubits. In particular, we notice that increasing the number of layers does not have a significant impact, at least at these scales, except perhaps in convergence speed of the training. It is likely however, that at larger scales, increased parameter numbers would be required to improve performance.

\subsection{Differing numbers of Boltzmann hidden nodes} \label{app_a_subapp:differing_number_hidden_nodes}

\begin{figure}[ht!]
    \centering
    \includegraphics[width=\columnwidth, height=0.35\columnwidth]{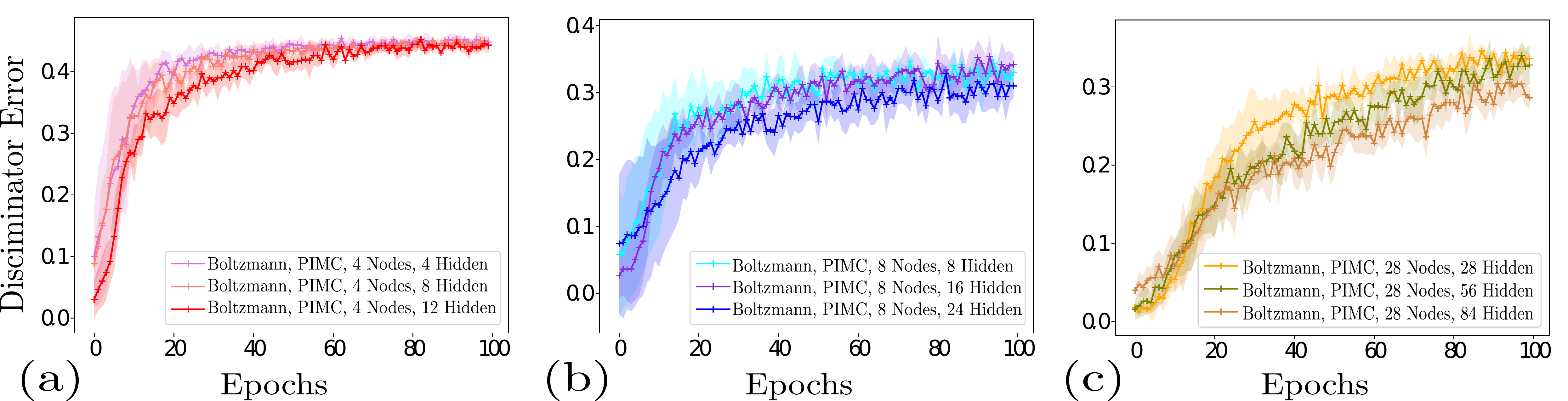}
    \caption{Increasing number of hidden nodes for RBMs with (a) $4$ (b) $8$ (c) $28$ visible nodes. Enlarging the hidden space for the RBM again did not impact significantly for these problem sizes and in particular would not give a performance boost to outperform the QCBM.}
    \label{fig:differing_hidden_nodes}
\end{figure}

We also demonstrate the effect of changing the number of hidden nodes in the Boltzmann machine in \Fref{fig:differing_hidden_nodes}, where we have $4, 8$ and $28$ visible nodes. Again, we observe that an increasing number of hidden nodes (and by extension, number of parameters) does not substantially affect the performance of the model, in fact it can hinder it, at least when training only biases of the Boltzmann machine. In particular, it does not substantially alter the final accuracy achieved by the model. We also noticed similar behavior when also training the weights of the Boltzmann machine.  

\subsection{Weight training of Boltzmann machine} \label{app_b_subapp:boltzmann_weight_training}

\begin{figure}[ht!]
    \centering
    \includegraphics[width=\columnwidth, height=0.35\columnwidth]{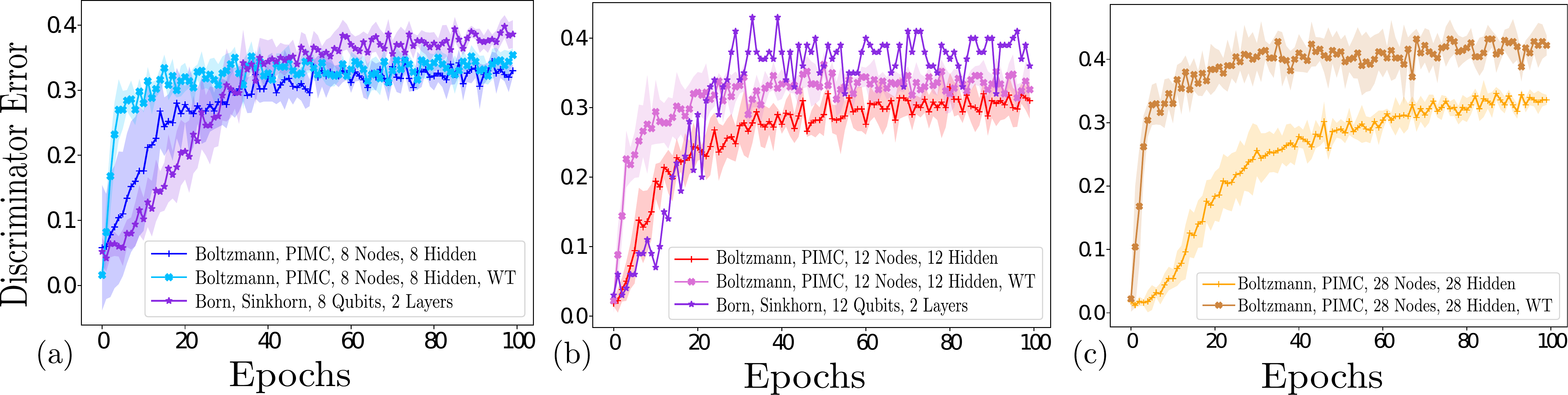}
    \caption{Weight training (WT) on the Boltzmann machine along with the node biases. We compare (a) $8$, (b) $12$ and (c) $28$ visible node RBMs along with the corresponding Born machine. The latter uses $4$ currency pairs, while the others use $2$, as in the text above. }
    \label{fig:weight_training}
\end{figure}

Finally, we compare the effect of weight training of the Boltzmann machine to training the bias terms alone. For the problem instances where the Boltzmann machine was able to converge to the best discriminator accuracy (i.e.\@ in the small problem instances), we find training the weights has the effect of increasing convergence speed, and also increased accuracy where training the biases only was insufficient to achieve high discriminator error. Interestingly, we note that the Born machine still outperforms the $8$ and $12$ visible node RBMs, even when the weights are also trained, and this does not seem to majorly affect the performance. However, training the weights does make a large difference for $28$ nodes, as seen in \Fref{fig:weight_training}(c), so again further investigation is needed in future work of this phenomenon.

\end{document}